\documentclass[prl,twocolumn,tightenlines,nofootinbib,nolongbibliography,aps,10pt]{revtex4-2}
\usepackage{bm,dcolumn,amsmath,graphicx,amsfonts,amssymb}

\usepackage{hyperref}
\usepackage{xcolor}
\definecolor{cite}{rgb}{0.,0.,0.9}   
\hypersetup{colorlinks,linkcolor={cite},citecolor={cite},urlcolor={cite}}

\usepackage[capitalise]{cleveref}
\usepackage{chemformula}
\usepackage{siunitx}

\renewcommand{\vec}[1]{\ensuremath{{\boldsymbol{#1}}}}
\newcommand{\abs}[1]{\ensuremath{\left |#1\right |}}
\newcommand{\bra}[1]{\ensuremath{\langle #1|}}	
\newcommand{\ket}[1]{\ensuremath{|#1\rangle}}	

\newcommand{\smallspace}{\rule{0pt}{2.5ex}}

\newcommand{\A}{\ensuremath{\mathcal{A}}} 

\begin{document}

\title{The hyperfine anomaly in mercury and test of the Moskowitz-Lombardi rule}

\author{J. Vandeleur}
\author{G. Sanamyan}
\author{B. M. Roberts}
\author{J. S. M. Ginges}
\affiliation{School of Mathematics and Physics, The University of Queensland, Brisbane QLD 4072, Australia}

\date{\today}

\begin{abstract}

We test the Moskowitz-Lombardi rule, originally formulated for mercury,
which gives a simple relation between the magnetic moment of an atomic nucleus and the effect of its radial distribution on the hyperfine structure -- the magnetic hyperfine anomaly or Bohr-Weisskopf (BW) effect. 
While the relation for the differential effect between isotopes may be completely determined experimentally, the value for the additive constant that is needed to give the BW effect for a single isotope has remained unverified.  
In this work, we determine the BW effect in H-like, singly-ionized, and neutral mercury isotopes from experimental muonic-$^{199}$Hg data 
together with differential anomalies and our atomic calculations. We check this result by directly extracting the BW effect from the measured hyperfine constant for $^{199}$Hg$^+$ using state-of-the-art atomic many-body calculations. From this we deduce an empirical value for the additive constant in the Moskowitz-Lombardi rule, which differs significantly from the values advocated previously.  
This result allows for increased precision in calculations of the hyperfine structure, and improved tests of atomic and nuclear theory.

\end{abstract}

\maketitle


High-precision atomic and molecular experiments provide unique, low-energy tests of the standard model of particle physics and searches for new particles and interactions~\cite{Ginges2004,Roberts2015,Safronova2018,Arrowsmith-Kron2024}. This includes studies of violations of fundamental symmetries (atomic parity violation~\cite{Khriplovich1991}, electric dipole moments~\cite{Khriplovich1997}), variation of fundamental constants~\cite{Dzuba1999}, and fifth-force searches~\cite{Berengut2018}. The interpretation of these experiments in terms of fundamental parameters relies on an accurate understanding of nuclear structure effects on the atom or molecule.  

As well as the finite distribution of nuclear charge, which significantly influences the electron wave functions and energies, there are other aspects of nuclear structure that must be understood and controlled in order for new-physics effects to be discernible. This includes deformation of the nuclear charge distribution, which must be accounted for in fifth-force searches in atomic clocks~\cite{Counts2020,Allehabi2021}, and which leads to orders-of-magnitude enhancement of CP-violating effects in electric dipole moment searches in diamagnetic systems through quadrupole~\cite{Flambaum1994} and octupole (pear-shaped)~\cite{Auerbach1996,Engel2000} deformation mechanisms. For systems with non-zero nuclear spin and therefore nuclear magnetic moments, the importance of an accurate understanding of the effect of the radial distribution of the nuclear magnetic moment on the hyperfine splitting for determination of nuclear magnetic moments, tests of nuclear structure, and tests of atomic theory in precision new-physics studies 
has been gaining recognition (see, e.g., Refs.~\cite{MP1995,Ginges2017,Konovalova2018,Konovalova2020,RobertsFr2020,Barzakh2020,Demidov2021,Roberts2021,Skripnikov2020,Roberts2021scr,Prosnyak2021,Skripnikov2022,Sanamyan2023,Lechner2023,Wilkins2023,Vandeleur2024}).

The effect of a finite distribution of the nuclear magnetic moment on the hyperfine structure is known as the Bohr-Weisskopf (BW) effect~\cite{Bohr1950,Bohr1951}. 
Direct calculation of the BW effect in a particular system requires knowledge of the distribution of nucleons and their contributions to the total magnetic moment~\cite{Fujita1975,Buttgenbach1984,Shabaev1994}. Microscopic nuclear calculations are not yet at the level where the BW effect may be reliably determined, and theoretical predictions for the effect vary significantly between models of nuclear magnetization~\cite{Senkov2002,Tomaselli2002,Ginges2017}.

A long-standing rule, proposed by Moskowitz and Lombardi \cite{Moskowitz1973}, relates the relative BW effect $\epsilon$ to the inverse of the magnitude of the nuclear magnetic moment $\mu$.   
This rule was originally formulated for mercury isotopes, based on empirical data for differential anomalies and nuclear magnetic moments, and expressed as a proportionate relation, $\epsilon \propto 1/|\mu|$. 
A theoretical analysis by Fujita and Arima~\cite{Fujita1975} led to the inclusion of an additive constant,  
and Moskowitz and Lombardi accepted this amendment in a later work that extended their rule to Au, Tl, and Ir~\cite{Moskowitz1982}. 
This ``empirical" rule has been used widely 
for the treatment of not only differential but also absolute anomalies. However, it is not universally valid, as has been demonstrated, for instance, in differential anomalies for Cd isotopes with nuclear spin $1/2^+$ \cite{Frommgen2015} and for the lanthanides Nd, Eu, and Gd~\cite{Persson2020}. Moreover, the value of the additive constant is experimentally inaccessible in most systems, and has remained largely untested. In particular, to our knowledge, until now there has been no empirical test of the constant for Hg.  

Mercury is a system of great interest in contemporary atomic and nuclear physics. The most stringent limit on a permanent electric dipole moment of an atom has been made for Hg, setting some of the tightest constraints on new sources of CP-violation~\cite{Graner2016}, while the interpretation is limited by the large theory uncertainty of the nuclear Schiff moment (up to several hundred percent)~\cite{Engel2013,Safronova2018}. 
Singly-ionized and neutral Hg are used as high-accuracy 
atomic clocks ~\cite{Berkeland1998,Diddams2001,Hachisu2008,Tyumenev2016,Schelfhout2022}, including in space-based applications~\cite{Burt:2021}, and the Hg nucleus is of interest for investigations of nuclear shape co-existence \cite{Sels2019, Marsh2018, Heyde2011}. These pursuits benefit from a better understanding of finite-nucleus effects and the hyperfine structure in this atom.

In this work, we determine empirical values for the BW effect in Hg isotopes and test the Moskowitz-Lombardi rule. We begin by extracting an updated experimental value for the BW effect in the 1$s_{1/2}$ state of muonic $^{199}$Hg, and we translate this into a BW effect for singly-ionized and neutral $^{199}$Hg. Furthermore, we perform state-of-the-art atomic many-body calculations for the magnetic hyperfine constant for $^{199}$Hg$^+$ and extract the BW effect directly from comparison with experiment. This latter result agrees with our muonic-deduced value, though with twice the uncertainty. From measured differential anomalies, we obtain the BW effect for a number of mercury isotopes.

Due to its close proximity to the nucleus, a muon in the 1$s$ state of a muonic atom is largely unscreened by atomic electrons. Therefore, this muonic atom may be treated like a hydrogenlike ion with a ``heavy" electron, with the muonic-atom wave functions found from the Dirac equation in the nuclear Coulomb field~\cite{Borie1982}. We take the distribution of nuclear charge to have the form of a Fermi distribution, with the 90\% to 10\% fall-off distance being \SI{2.3}{\femto\metre}, and root-mean-square charge radii taken from Ref.~\cite{Angeli2013}. 

The relativistic operator for the magnetic hyperfine interaction is 
\begin{equation} 
    h_\text{hfs} = \alpha \vec{\mu} \cdot \left( \vec{r} \times \vec{\alpha} \right)F(r) /r^3 \, ,
    \label{eqn:HFSoperator}
\end{equation}
where $\vec{\mu}=\mu\vec{I}/I$ is the nuclear magnetic moment, $\vec{I}$ is the nuclear spin, $\vec{\alpha}$ is the Dirac matrix, $\vec{r}$ is the polar radial vector, and $F(r)$ accounts for the finite radial distribution of the nuclear magnetic moment ($F(r) = 1$ for the point-like case). Here we use atomic units $\hbar = m_e = \abs{e} = c \alpha = 1$.

The magnetic hyperfine structure is often quantified through the hyperfine constant, $\A$, which is proportional to the expectation value of the operator~\cref{eqn:HFSoperator}, $\langle \varphi |h_{\rm hfs}|\varphi\rangle =\mathcal{A}\langle\vec{I}\cdot \vec{J}\rangle$, where $\vec{J}$ is the total electronic angular momentum. It is convenient to express the hyperfine constant in terms of distinct contributions, 
\begin{equation}
\label{eq:contributions}
\A=\A_0+\A_{\rm BW}+\A_{\rm QED} \, .
\end{equation}
There are also higher-order contributions, including the nuclear polarization; however, these are small~\cite{Vandeleur2024} and captured by the uncertainties of the considered terms.   
The leading term $\mathcal{A}_0$ includes account of the finite-nuclear-charge distribution -- the Breit-Rosenthal effect~\cite{Rosenthal1932,Crawford1949}. The BW effect, from the finite distribution of the magnetic moment, may be expressed as a relative correction $\epsilon$,
\begin{align}
    \A_0+\A_{\rm BW}&=\A_{0}(1+\epsilon)\, . 
    \label{eqn:A0point}
\end{align}
Finally, a correction to the hyperfine constant arises from quantum electrodynamics (QED), $\mathcal{A}_{\rm QED}$.  
For muonic atoms, the vacuum polarization contributes overwhelmingly to this correction and the more complicated self-energy may be omitted at the considered level of precision~\cite{Borie1982,Elizarov2006,Sanamyan2023}. The electric- and magnetic-loop vacuum polarization corrections may be evaluated in the Uehling approximation~\cite{Uehling1935} (see, e.g., Ref.~\cite{Elizarov2006,Volotka2008}), as we have done in this work.

\begin{table}
    \centering
    \caption{Extracted BW effects $\A_{\rm BW,\,exp}$ in $\mu$-$^{199}$Hg $1s$ and $^{199}$Hg$^+$ $6s$ from measured hyperfine constants $\A_{\rm exp}$ using our point-nucleus theory results $\A_0$ and QED values $\A_{\rm QED}$.}
    \label{tbl:HgBreakDown}
    \begin{ruledtabular}
    \begin{tabular}{ccccc}
        & $\A_{\rm exp}$ & $\A_0$ & $\A_{\rm QED}$ & $\A_{\rm BW,\,exp}$ \\
        \hline
       $\mu$-Hg\ (keV) &0.47(12)\footnotemark[1] & 1.449(2) & 0.007(4) & -0.99(12) \\
         Hg$^+$\ (MHz) & 40507.347...\footnotemark[2] & 42170(420) & -240(48)\footnotemark[3] & -1420(420)
    \end{tabular}
    \end{ruledtabular}
    \footnotetext[1]{Ref.~\cite{Buttgenbach1984};  $^{\rm b}$ Ref.~\cite{Berkeland1998}; 
    $^{\rm c}$ Interpolated from results of Ref.~\cite{Ginges2017}.}
\end{table}

Even without knowledge of the radial distribution of the magnetic dipole moment within the nucleus, it is possible to extract its effect on the hyperfine constant of muonic atoms from a comparison of theory evaluations for $\mathcal{A}_0$, $\mathcal{A}_{\rm QED}$, and the measured value $\mathcal{A}_{\rm exp}$; see Eq.~(\ref{eq:contributions}). The results of our calculations and the measured value of the hyperfine constant are shown in the first row of \cref{tbl:HgBreakDown} for the $1s$ state of muonic $^{199}$Hg, along with the extracted value for the BW effect. We find this latter value to be $\mathcal{A}_{\rm BW,\,exp}=-0.99(12)\,{\rm keV}$, which is about two-thirds the size of the leading contribution $\mathcal{A}_0$ and of opposite sign. The uncertainty is dominated by that of the measurement. Expressed as a relative correction, the extracted BW effect corresponds to $\epsilon_{\rm exp} = -68(8)\%$, which is the same value determined previously~\cite{Buttgenbach1984}, while the theory contributions are slightly different. This empirically-deduced value is smaller than those from nuclear theory. For example, in the simple nuclear single-particle model -- with the value for the nuclear magnetic moment ascribed to the unpaired neutron~\cite{Shabaev1994,Volotka2008} -- we obtain $-85\%$, while microscopic nuclear theory calculations~\cite{Fujita1975} give $-81.1\%$ or $-83.8\%$, depending on the chosen nuclear parameters.  

The obtained empirical BW effect may be translated from muonic to atomic Hg by following the two-step method set out in Ref.~\cite{Sanamyan2023}. In the first step, the effect is translated to that for the H-like ion, as originally proposed in Ref.~\cite{Elizarov2006}; and in the second step, the H-like result is translated to that for a many-electron system by introducing electronic screening factors~\cite{Roberts2021scr} (see also Ref.~\cite{Skripnikov2020}). In the first step, it is important to capture a reasonable uncertainty for the  nuclear model-dependence. We do this within the single-particle model where the magnetic moment's radial distribution $F(r)$ is found from spin and orbital g-factors and explicit forms for the radial wave function of the unpaired nucleon~\cite{Elizarov2006}. Following Ref.~\cite{Elizarov2006}, we consider five distinct polynomial shapes for the nucleon wave function which vary as $r^n$ and $(r_m-r)^n$, for $n=0,1,2$.
For each of these, the effective magnetic radius $r_m$ is varied until the empirical BW effect is reproduced in muonic Hg, and then the same $F(r)$ is applied to the H-like ion to find the corresponding BW effect in that system. The range in values from the different distributions gives an estimate of the nuclear model dependence. Using this procedure, we obtain a relative BW effect of $-2.39(45)\%$ in $^{199}$Hg$^{79+}$. The largest part of the uncertainty originates from the muonic-Hg experiment, with a sizeable part from the nuclear model dependence.

\begin{table}[tb]
    \centering
    \caption{Electronic screening factors $x_{\rm scr}$ and BW effects $\epsilon$ (in \%) for $\mu$-atom, H-like, singly-ionized, and neutral $^{199}$Hg found from $\mu$-$^{199}$Hg experiment, alongside ML and FA values.}
    \label{tbl:BWscr}
    \begin{ruledtabular}
    \begin{tabular}{lcccc}
& $\mu$-$^{199}$Hg & $^{199}$Hg$^{79+}$ & $^{199}$Hg$^+$ & $^{199}$Hg \\
    \hline
$x_{\rm scr}$ & & 1 & 1.086(3) & 0.980(5) 
\\ 
$\epsilon_{\rm exp}$ &-68(8) &-2.39(45)
& -2.59(49),\, -3.4(10)\footnotemark[1] 
& -2.34(44) \\
$\epsilon^{\rm ML}$~\cite{Moskowitz1973} & & -2.1\footnotemark[2]&-2.2\footnotemark[3] & -2.0 \\
$\epsilon^{\rm FA}$~\cite{Fujita1975} & &-3.5\footnotemark[2] & -3.8\footnotemark[3] & -3.4 \\
    \end{tabular}
    \end{ruledtabular}
\footnotetext[1]{From direct extraction in Hg$^+$, see \cref{tbl:HgBreakDown}.}
\footnotetext[2]{Obtained from Hg$^+$ value using screening factor in first row.}
\footnotetext[3]{Obtained from Hg value using screening factor, \cref{eq:singleScreen}.}      
\end{table}

Since the BW effect originates at small distances from the center of the nucleus where only $s$ and $p_{1/2}$ orbitals penetrate, and such orbitals are proportional within the nucleus to a good approximation, all the information about the BW effect is contained within the H-like result~\cite{Shabaev2001,Skripnikov2020,Roberts2021scr}. One may then introduce electronic screening factors, $x_{\rm scr}=\epsilon^{\rm atom}/\epsilon^{\rm H-like}$, and determine the effect in many-electron atoms, independent of the nuclear model~\cite{Roberts2021scr}. We have performed relativistic atomic many-body calculations for singly-ionized Hg to obtain this factor, $x_{\rm scr}=\epsilon(^{199}{\rm Hg}^+\, 6s)/\epsilon(^{199}{\rm Hg}^{79+}\, 1s)$, which we present in \cref{tbl:BWscr}. The screening calculations were performed at the level of random phase approximation with exchange. The contribution of higher-order correlations to the screening effect is very small; see Refs.~\cite{Roberts2021,Roberts2021scr} for demonstration of this and details of the method. Applying the screening factor to the muonic-atom-derived H-like value, we obtain $\epsilon_{\rm exp} = -2.59(49)\%$ for the BW effect in the ground state of $^{199}$Hg$^+$. 

For neutral Hg, the electronic screening calculations are more complicated, with a correspondingly larger uncertainty. Rather than finding this from theory, we obtain a screening factor that relates the BW effects in neutral and singly-ionized Hg from the ratio of measured differential anomalies for isotopes $199$ and $201$~\cite{Persson2023}, 
\begin{equation}
    \frac{{^{199}\Delta^{201}} (6s6p\  {^3{\rm P}}_{1})}{{^{199}\Delta^{201}}(6s\ {^2{\rm S}}_{1/2})} = 0.902(4) \ . \label{eq:singleScreen}
\end{equation}
 The differential anomaly $^{1}\Delta^{2}$ between isotopes $1$ and $2$ is given by~\cite{Persson2023}
\begin{equation}
    ^{1}\Delta^{2} = \left( \A_1/\A_2 \right) \left( \mu_2 I_1/ \mu_1 I_2\right) - 1 \approx {^{1}\delta^{2}} + \epsilon^{(1)} - \epsilon^{(2)},
    \label{eqn:HFanomaly}
\end{equation}
where $^{1}\delta^{2} = \delta^{(1)} - \delta^{(2)}$ is the differential Breit-Rosenthal effect. 
We have checked numerically that $^{199}\delta^{201}$ is very small and may be omitted in the ratio (\ref{eq:singleScreen}), and so we can take this to be the ratio of BW effects. Therefore, we find the corresponding BW effect in neutral $^{199}$Hg, originally derived from the muonic-atom experiment, which we present in the final column of Table~\ref{tbl:BWscr}. Furthermore, as there is experimental data for differential anomalies for a number of Hg isotopes, the BW effect for these isotopes may be found. For those isotopes with sufficient spin, any electric quadrupole contribution to the hyperfine splitting has already been removed to isolate the magnetic hyperfine constant and infer the differential anomaly; see Ref.~\cite{Reimann1973}. 
The isotopes with their nuclear parameters, differential anomalies, and BW effects for Hg and Hg$^+$ are shown in Table~\ref{tbl:MLPlot}. 
The differential anomalies for the ions are determined from the neutral-atom data~\cite{Persson2023} using Eq.~\eqref{eq:singleScreen}.
The small differential Breit-Rosenthal effects are included in the computation of $\epsilon$. Note that the uncertainty in the BW effect for each of the isotopes for Hg and Hg$^+$ has been propagated from the uncertainty in the value for $^{199}$Hg$^{79+}$, itself determined from that in $\mu$-$^{199}$Hg arising largely from the muonic-atom experiment.  

The size of the BW effect in $^{199}$Hg$^+$ $6s$ is quite large compared to the typical size of such effects in electronic systems (several 0.1\%), and it is significantly larger than the accuracy of state-of-the-art atomic many-body calculations for heavy, single-valence electron atoms~\cite{Ginges2017,Grunefeld2019,Roberts2022}. Therefore, as an independent check of the result obtained from muonic Hg, we perform high-precision atomic calculations for the hyperfine constant for $^{199}$Hg$^+$ using the all-orders correlation potential approach~\cite{Dzuba1988} to find the BW effect directly from comparison with experiment. Details of the methods applied to hyperfine structure may be found, e.g., in Refs.~\cite{Ginges2017,Grunefeld2019,RobertsFr2020}. 
At the relativistic Hartree-Fock level with core polarization (random phase approximation with exchange) our result is 38360\,MHz. Including all-orders valence-core correlations, energy-fitting, structural radiation, and normalization gives 42174\,MHz. We estimate the size of QED corrections from interpolation of the relative corrections for Ba$^+$ and Ra$^+$~\cite{Ginges2017}, assigning a 
20\% uncertainty. From an analysis of the accuracy of our results, which includes benchmarking of the calculated energies against experimental values and consideration of the size of correlations and stability of the results, we estimate the uncertainty to be $1\%$ for the point-magnetization value $\mathcal{A}_0$. Our results for the hyperfine constant for $^{199}$Hg$^+$ are shown in the second row of \cref{tbl:HgBreakDown}. From comparison with the experimental value for the hyperfine constant for the same ion, we directly extract the BW effect $\epsilon = -3.4(10)\%$, which is in agreement with our value obtained from the muonic-atom experiment [-2.59(49)\%], though with twice the uncertainty.

\begin{table*}[tbh]
    \centering
    \caption{Nuclear spins $I^\pi$, rms radii, moments $\mu$, differential anomalies $\Delta$, and relative BW effects $\epsilon_{\rm exp}$ in Hg$^+$ and Hg isotopes. BW effect in $^{199}$Hg$^+$ is found from $\mu$-$^{199}$Hg and translated to other isotopes and neutral Hg using measured differential anomalies~\cite{Persson2023} and the screening factor~\cref{eq:singleScreen}.}
    \label{tbl:MLPlot}
    \begin{ruledtabular}
    \begin{tabular}{@{}lclccccc@{}}
    &&&&\multicolumn{2}{c}{Hg ($6s6p$ $^3$P$_1$)}&\multicolumn{2}{c}{Hg$^+$ ($6s$ $^2$S$_{1/2}$)}\\
    \cline{5-6}\cline{7-8}\smallspace
    $A$   &  $I^\pi$  &  $r_{\rm rms}$ (fm) \cite{Angeli2013}   &  $\mu \ (\mu_{\rm N})$ \cite{Stone2019}  &  $^A\Delta^{199}$ \cite{Persson2023} &  $-\epsilon_{\rm exp} \ (\%)$  &  $^A\Delta^{199}$  &  $-\epsilon_{\rm exp} \ (\%)$\\ \hline
    $193$          &  $3 /2^-$   &  5.4238(35)  &  -0.6251(8)  &  $0.61(3)$  &  1.75(44) &   $0.68(4)$   &  1.94(49)\\
    $193^{\rm m}$  &  $13 /2^+$  &  5.4264(31)\footnotemark[1]  &  -1.0543(12)  &  $1.0552(13)$  &  1.30(44)  &   $1.17(4)$  &  1.44(49) \\
    $195$          &  $1 /2^-$   &  5.4345(32)  &  0.5393(6)   &  $0.1470(9)$  &  2.20(44)  &  $0.163(5)$  &  2.44(49)\\
    $195^{\rm m}$  &  $13 /2^+$  &  5.4342(31)\footnotemark[1]  &  -1.0405(12)  &  $1.038(3)$  &  1.31(44)  &   $1.15(4)$  &  1.45(49) \\  
    $197$          &  $1 /2^-$   &  5.4414(31)  &  0.5253(6)  &  $0.0778(7)$  &  2.26(44)  &  $0.086(3)$  &  2.51(49)\\
    $197^{\rm m}$  &  $13 /2^+$  &  5.4424(31)\footnotemark[1]  &  -1.0236(12)  &  $1.021(3)$  &  1.32(44)  &   $1.13(3)$  &  1.46(49)\\
    $199$          &  $1 /2^-$   &  5.4474(31)  &  0.5039(6)  & 0 &  2.34(44)  & 0 &  2.59(49) \\
    $199^{\rm m}$  &  $13 /2^+$  &  5.4520(31)\footnotemark[1]   &  -1.0107(12)  &  $0.960(9)$  &  1.37(44)  &  $1.07(3)$  &   1.52(49)\\
    $201$          &  $3 /2^-$   &  5.4581(32)  &  -0.5580(7)  &  $0.1467(6)$  &  2.18(44)  &  $0.16257(5)$\footnotemark[2]  &  2.42(49)\\ 
    $203$          &  $5 /2^-$   &  5.4679(35)  &  0.8456(10)  &  $0.796(16)$  &  1.52(44)  &   $0.88(3)$   &  1.68(49)\\
    \end{tabular}
    \end{ruledtabular}
    \footnotetext[1]{From $\delta\langle r^2 \rangle^{198,A}$~\cite{Ulm1986} and $r_{\rm rms}^{(198)}=5.4463(31)$\,fm~\cite{Angeli2013};
    $^{\rm b}$ Direct measurement~\cite{Burt2009}.}
\end{table*}

In the original work of Moskowitz and Lombardi~\cite{Moskowitz1973}, they proposed $\epsilon = \alpha_{\rm ML}/\abs{\mu}$, with $\alpha_{\rm ML} = 1.0\times 10^{-2}\mu_{\rm N}$, where $\mu_N$ is the nuclear magneton.  
This was extended by Fujita and Arima~\cite{Fujita1975} to the form
\begin{equation}
    \epsilon = c - \frac{\alpha}{\abs{\mu}} \ = \ \epsilon^{(0)} + \epsilon^{(1)}  ,  
    \label{eq:ML}
\end{equation}
and for Hg they proposed the coefficients
\begin{equation}
c_{\rm FA} = -0.01 , \ \alpha_{\rm FA} = 1.2\times10^{-2} \mu_{\rm N} \, , \label{eq:FA_params}
\end{equation}
which reproduce the differential anomalies with an uncertainty $10-20\%$. 
The expression~(\ref{eq:ML}) has a basis in nuclear and atomic theory~\cite{Fujita1975},  
\begin{gather}
    \epsilon^{(0)} = -0.62b^{(\kappa)}\langle R^2 \rangle \, ,
    \label{eq:FA0} \\
    \epsilon^{(1)} = -\frac{0.38}{\mu}b^{(\kappa)}\left\langle R^2\right\rangle\bra{II}\sum_{i=1}^Ag_s^{(i)}\Sigma_i^{(1)}\ket{II} \label{eq:FA1}\, .
\end{gather}
Here, $\langle R^2\rangle$ is the lowest radial magnetization moment, the factor $b^{({\kappa})}$ is determined from the electronic wave functions in the nuclear vicinity for relativistic angular momentum quantum number $\kappa$, $g_s^{(i)}$ is the spin g-factor of the nucleon $i$, and $\Sigma_i^{(1)}=s_i+\sqrt{2} (sY^{(2)})^{(1)}_i$ is a nuclear operator comprised of the nucleon spin $s_i$ and spin-asymmetry term, with the latter arising from the tensor product of the spin and spherical harmonic $Y^{(2)}$; see Refs.~\cite{Bohr1950,Bohr1951,Fujita1975,Shabaev1994} for details. 
Along with some other simplifications, the expressions~(\ref{eq:FA0}) and (\ref{eq:FA1}) are based on the factorization of the nuclear wave function into radial and angular parts, which may be valid as long as the BW effect is dominated by unpaired nucleons or that the distribution of core nucleons is similar to the unpaired ones~\cite{Fujita1975}. The former condition may be satisfied close to the magic shell closures, and indeed it is in the region of Pb that the Moskowitz-Lombardi rule has been observed to hold~\cite{Persson2020}. 

\begin{figure}[htb]
\centering
\includegraphics[width=1.\linewidth]{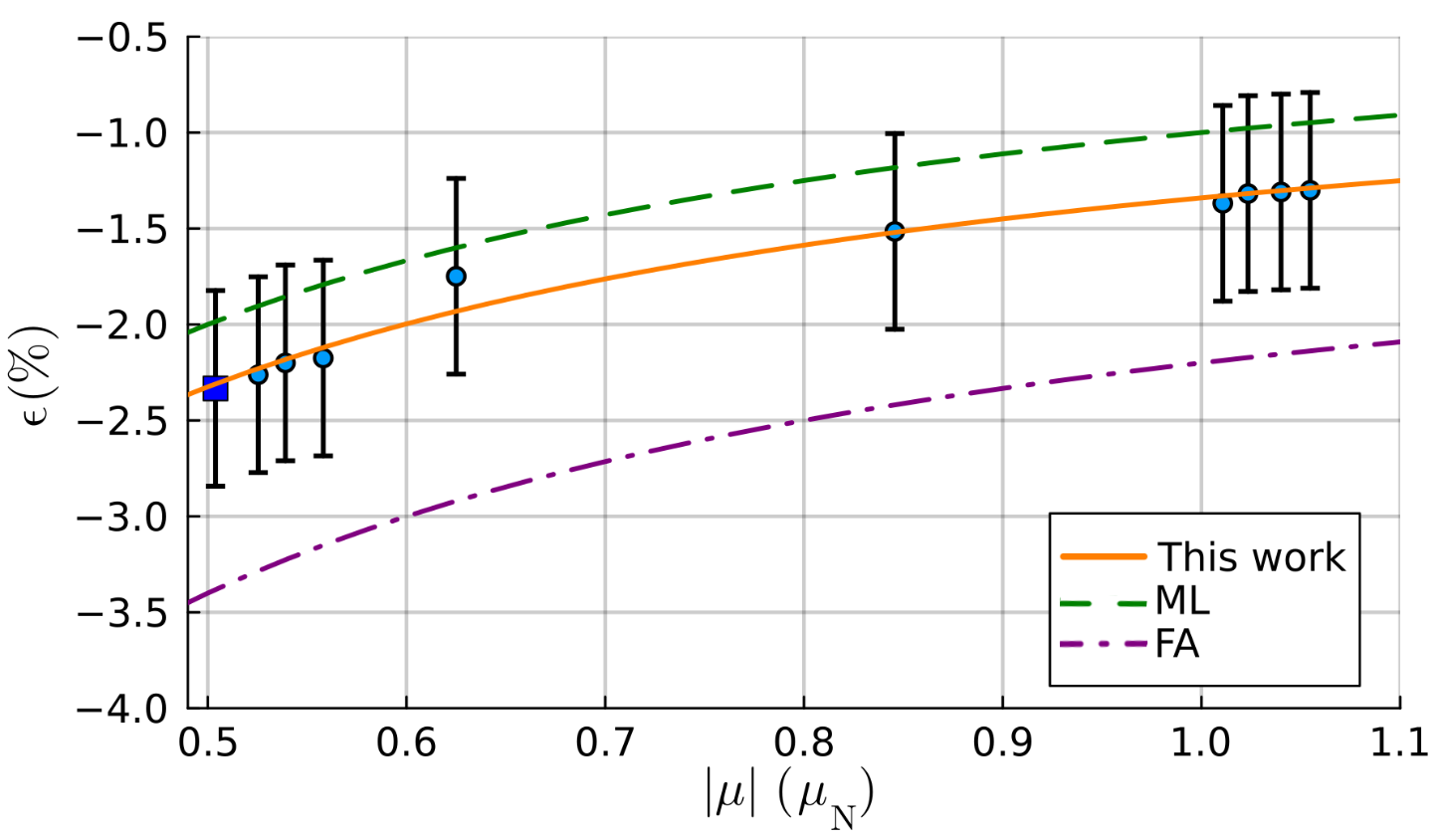}
\caption{Fit of the Moskowitz-Lombardi rule to the relative BW effects $\epsilon$ for neutral Hg isotopes (\cref{tbl:MLPlot}), together with the models of Moskowitz and Lombardi (ML) ~\cite{Moskowitz1973} and Fujita and Arima (FA) \cite{Fujita1975}. The result for $^{199}$Hg is indicated by a square.}
\label{fig:MLPlot}
\end{figure}

Our extracted BW effects may be compared to those of Moskowitz-Lombardi (ML) and Fujita-Arima (FA). For $^{199}$Hg, we obtain $\epsilon_{\rm exp} = -2.34(44)\%$, while the ML and FA values are $-2.0\%$ and $-3.4\%$, respectively; see Table~\ref{tbl:BWscr}. Our result agrees with the ML value, though differs substantially from that of FA. 
Our determination of the BW effect for Hg isotopes allows us to find parameters $c$  and $\alpha$ by performing a fit of Eq.~(\ref{eq:ML}) to the obtained data in Table~\ref{tbl:MLPlot}, 
\begin{gather}
 c_{\rm exp}=-3.6(8)\times10^{-3} , \ \alpha_{\rm exp} = 0.98(5)\times10^{-2}\mu_{\rm N}.
 \label{eq:Hg_params} 
\end{gather}
For singly-ionized Hg, we find the parameters 
$c_{\rm exp}=-4.0(9)\times10^{-3}$ and  $\alpha_{\rm exp} = 1.09(6)\times10^{-2}\mu_{\rm N}$. 
In \cref{fig:MLPlot} we plot the obtained BW effects for isotopes of neutral Hg, including the fit to this data, alongside the ML and FA curves. While the empirical values determined in this work are consistent with the ML data, it is seen that there is significant deviation from the theory-based values of FA. In particular, our results indicate that the additive term $c$ is smaller than suggested in this latter work. The deviations arise from omitted many-body nuclear effects in the theory analysis of Ref.~\cite{Fujita1975}, and may be used to inform nuclear microscopic models. 

We have determined empirical values for the Bohr-Weisskopf effects in ten isotopes and isomers of neutral and singly-ionized mercury. These have been found from the measured hyperfine splitting in the $1s$ state of muonic-$^{199}$Hg, in combination with our atomic calculations and measured differential anomalies. Until now, only the differential anomalies have been known for Hg. Determination of the absolute anomalies has allowed us to test the well-known Moskowitz-Lombardi rule, and provide more accurate coefficients in its implementation. 
The uncertainty in our result is dominated by that of the original muonic-Hg experiment. Contemporary experiments in muonic atoms~\cite{Antognini2020,Wauters2021} or in H-like or few-electron highly-charged ions~\cite{Ullmann2017,Dickopf2024} could reduce this significantly. 
Through improved knowledge of nuclear magnetization distribution effects, we have provided much-needed information for precision atomic calculations of the hyperfine structure, tests of atomic theory in the nuclear region through comparison of theoretical and experimental hyperfine constants, and tests of nuclear structure theory in Hg. 
Through these avenues, our results may be used to increase the discovery potential of low-energy searches for new physics beyond the standard model.

\acknowledgments
This work was supported by the Australian Government through an Australian Research Council (ARC) DECRA Fellowship DE210101026, ARC Future Fellowship FT170100452, and ARC Discovery Project DP230101685.

\bibliography{library}

\begin{thebibliography}{71}%
\makeatletter
\providecommand \@ifxundefined [1]{%
 \@ifx{#1\undefined}
}%
\providecommand \@ifnum [1]{%
 \ifnum #1\expandafter \@firstoftwo
 \else \expandafter \@secondoftwo
 \fi
}%
\providecommand \@ifx [1]{%
 \ifx #1\expandafter \@firstoftwo
 \else \expandafter \@secondoftwo
 \fi
}%
\providecommand \natexlab [1]{#1}%
\providecommand \enquote  [1]{``#1''}%
\providecommand \bibnamefont  [1]{#1}%
\providecommand \bibfnamefont [1]{#1}%
\providecommand \citenamefont [1]{#1}%
\providecommand \href@noop [0]{\@secondoftwo}%
\providecommand \href [0]{\begingroup \@sanitize@url \@href}%
\providecommand \@href[1]{\@@startlink{#1}\@@href}%
\providecommand \@@href[1]{\endgroup#1\@@endlink}%
\providecommand \@sanitize@url [0]{\catcode `\\12\catcode `\$12\catcode
  `\&12\catcode `\#12\catcode `\^12\catcode `\_12\catcode `\%12\relax}%
\providecommand \@@startlink[1]{}%
\providecommand \@@endlink[0]{}%
\providecommand \url  [0]{\begingroup\@sanitize@url \@url }%
\providecommand \@url [1]{\endgroup\@href {#1}{\urlprefix }}%
\providecommand \urlprefix  [0]{URL }%
\providecommand \Eprint [0]{\href }%
\providecommand \doibase [0]{https://doi.org/}%
\providecommand \selectlanguage [0]{\@gobble}%
\providecommand \bibinfo  [0]{\@secondoftwo}%
\providecommand \bibfield  [0]{\@secondoftwo}%
\providecommand \translation [1]{[#1]}%
\providecommand \BibitemOpen [0]{}%
\providecommand \bibitemStop [0]{}%
\providecommand \bibitemNoStop [0]{.\EOS\space}%
\providecommand \EOS [0]{\spacefactor3000\relax}%
\providecommand \BibitemShut  [1]{\csname bibitem#1\endcsname}%
\let\auto@bib@innerbib\@empty
\bibitem [{\citenamefont {Ginges}\ and\ \citenamefont
  {Flambaum}(2004)}]{Ginges2004}%
  \BibitemOpen
  \bibfield  {author} {\bibinfo {author} {\bibfnamefont {J.~S.~M.}\
  \bibnamefont {Ginges}}\ and\ \bibinfo {author} {\bibfnamefont {V.~V.}\
  \bibnamefont {Flambaum}},\ }\href
  {https://doi.org/10.1016/j.physrep.2004.03.005} {\bibfield  {journal}
  {\bibinfo  {journal} {Phys. Rep.}\ }\textbf {\bibinfo {volume} {397}},\
  \bibinfo {pages} {63} (\bibinfo {year} {2004})}\BibitemShut {NoStop}%
\bibitem [{\citenamefont {Roberts}\ \emph {et~al.}(2015)\citenamefont
  {Roberts}, \citenamefont {Dzuba},\ and\ \citenamefont
  {Flambaum}}]{Roberts2015}%
  \BibitemOpen
  \bibfield  {author} {\bibinfo {author} {\bibfnamefont {B.~M.}\ \bibnamefont
  {Roberts}}, \bibinfo {author} {\bibfnamefont {V.~A.}\ \bibnamefont {Dzuba}},\
  and\ \bibinfo {author} {\bibfnamefont {V.~V.}\ \bibnamefont {Flambaum}},\
  }\href {https://doi.org/10.1146/annurev-nucl-102014-022331} {\bibfield
  {journal} {\bibinfo  {journal} {Annu. Rev. Nucl. Part. Sci.}\ }\textbf
  {\bibinfo {volume} {65}},\ \bibinfo {pages} {63} (\bibinfo {year}
  {2015})}\BibitemShut {NoStop}%
\bibitem [{\citenamefont {Safronova}\ \emph {et~al.}(2018)\citenamefont
  {Safronova}, \citenamefont {Budker}, \citenamefont {DeMille}, \citenamefont
  {Kimball}, \citenamefont {Derevianko},\ and\ \citenamefont
  {Clark}}]{Safronova2018}%
  \BibitemOpen
  \bibfield  {author} {\bibinfo {author} {\bibfnamefont {M.}~\bibnamefont
  {Safronova}}, \bibinfo {author} {\bibfnamefont {D.}~\bibnamefont {Budker}},
  \bibinfo {author} {\bibfnamefont {D.}~\bibnamefont {DeMille}}, \bibinfo
  {author} {\bibfnamefont {D.~F.~J.}\ \bibnamefont {Kimball}}, \bibinfo
  {author} {\bibfnamefont {A.}~\bibnamefont {Derevianko}},\ and\ \bibinfo
  {author} {\bibfnamefont {C.~W.}\ \bibnamefont {Clark}},\ }\href
  {https://doi.org/10.1103/revmodphys.90.025008} {\bibfield  {journal}
  {\bibinfo  {journal} {Rev. of Mod. Phys.}\ }\textbf {\bibinfo {volume}
  {90}},\ \bibinfo {pages} {025008} (\bibinfo {year} {2018})}\BibitemShut
  {NoStop}%
\bibitem [{\citenamefont {Arrowsmith-Kron}\ \emph {et~al.}(2024)\citenamefont
  {Arrowsmith-Kron}, \citenamefont {Athanasakis-Kaklamanakis}, \citenamefont
  {Au}, \citenamefont {Ballof}, \citenamefont {Berger}, \citenamefont
  {Borschevsky}, \citenamefont {Breier}, \citenamefont {Buchinger},
  \citenamefont {Budker}, \citenamefont {Caldwell}, \citenamefont {Charles},
  \citenamefont {Dattani}, \citenamefont {de~Groote}, \citenamefont {DeMille},
  \citenamefont {Dickel}, \citenamefont {Dobaczewski}, \citenamefont
  {Düllmann}, \citenamefont {Eliav}, \citenamefont {Engel}, \citenamefont
  {Fan}, \citenamefont {Flambaum}, \citenamefont {Flanagan}, \citenamefont
  {Gaiser}, \citenamefont {Ruiz}, \citenamefont {Gaul}, \citenamefont {Giesen},
  \citenamefont {Ginges}, \citenamefont {Gottberg}, \citenamefont {Gwinner},
  \citenamefont {Heinke}, \citenamefont {Hoekstra}, \citenamefont {Holt},
  \citenamefont {Hutzler}, \citenamefont {Jayich}, \citenamefont {Karthein},
  \citenamefont {Leach}, \citenamefont {Madison}, \citenamefont
  {Malbrunot-Ettenauer}, \citenamefont {Miyagi}, \citenamefont {Moore},
  \citenamefont {Moroch}, \citenamefont {Navratil}, \citenamefont {Nazarewicz},
  \citenamefont {Neyens}, \citenamefont {Norrgard}, \citenamefont {Nusgart},
  \citenamefont {Pašteka}, \citenamefont {Petrov}, \citenamefont {Plaß},
  \citenamefont {Ready}, \citenamefont {Reiter}, \citenamefont {Reponen},
  \citenamefont {Rothe}, \citenamefont {Safronova}, \citenamefont
  {Scheidenerger}, \citenamefont {Shindler}, \citenamefont {Singh},
  \citenamefont {Skripnikov}, \citenamefont {Titov}, \citenamefont {Udrescu},
  \citenamefont {Wilkins},\ and\ \citenamefont {Yang}}]{Arrowsmith-Kron2024}%
  \BibitemOpen
  \bibfield  {author} {\bibinfo {author} {\bibfnamefont {G.}~\bibnamefont
  {Arrowsmith-Kron}}, \bibinfo {author} {\bibfnamefont {M.}~\bibnamefont
  {Athanasakis-Kaklamanakis}}, \bibinfo {author} {\bibfnamefont
  {M.}~\bibnamefont {Au}}, \bibinfo {author} {\bibfnamefont {J.}~\bibnamefont
  {Ballof}}, \bibinfo {author} {\bibfnamefont {R.}~\bibnamefont {Berger}},
  \bibinfo {author} {\bibfnamefont {A.}~\bibnamefont {Borschevsky}}, \bibinfo
  {author} {\bibfnamefont {A.~A.}\ \bibnamefont {Breier}}, \bibinfo {author}
  {\bibfnamefont {F.}~\bibnamefont {Buchinger}}, \bibinfo {author}
  {\bibfnamefont {D.}~\bibnamefont {Budker}}, \bibinfo {author} {\bibfnamefont
  {L.}~\bibnamefont {Caldwell}}, \bibinfo {author} {\bibfnamefont
  {C.}~\bibnamefont {Charles}}, \bibinfo {author} {\bibfnamefont
  {N.}~\bibnamefont {Dattani}}, \bibinfo {author} {\bibfnamefont {R.~P.}\
  \bibnamefont {de~Groote}}, \bibinfo {author} {\bibfnamefont {D.}~\bibnamefont
  {DeMille}}, \bibinfo {author} {\bibfnamefont {T.}~\bibnamefont {Dickel}},
  \bibinfo {author} {\bibfnamefont {J.}~\bibnamefont {Dobaczewski}}, \bibinfo
  {author} {\bibfnamefont {C.~E.}\ \bibnamefont {Düllmann}}, \bibinfo {author}
  {\bibfnamefont {E.}~\bibnamefont {Eliav}}, \bibinfo {author} {\bibfnamefont
  {J.}~\bibnamefont {Engel}}, \bibinfo {author} {\bibfnamefont
  {M.}~\bibnamefont {Fan}}, \bibinfo {author} {\bibfnamefont {V.}~\bibnamefont
  {Flambaum}}, \bibinfo {author} {\bibfnamefont {K.~T.}\ \bibnamefont
  {Flanagan}}, \bibinfo {author} {\bibfnamefont {A.~N.}\ \bibnamefont
  {Gaiser}}, \bibinfo {author} {\bibfnamefont {R.~F.~G.}\ \bibnamefont {Ruiz}},
  \bibinfo {author} {\bibfnamefont {K.}~\bibnamefont {Gaul}}, \bibinfo {author}
  {\bibfnamefont {T.~F.}\ \bibnamefont {Giesen}}, \bibinfo {author}
  {\bibfnamefont {J.~S.~M.}\ \bibnamefont {Ginges}}, \bibinfo {author}
  {\bibfnamefont {A.}~\bibnamefont {Gottberg}}, \bibinfo {author}
  {\bibfnamefont {G.}~\bibnamefont {Gwinner}}, \bibinfo {author} {\bibfnamefont
  {R.}~\bibnamefont {Heinke}}, \bibinfo {author} {\bibfnamefont
  {S.}~\bibnamefont {Hoekstra}}, \bibinfo {author} {\bibfnamefont {J.~D.}\
  \bibnamefont {Holt}}, \bibinfo {author} {\bibfnamefont {N.~R.}\ \bibnamefont
  {Hutzler}}, \bibinfo {author} {\bibfnamefont {A.}~\bibnamefont {Jayich}},
  \bibinfo {author} {\bibfnamefont {J.}~\bibnamefont {Karthein}}, \bibinfo
  {author} {\bibfnamefont {K.~G.}\ \bibnamefont {Leach}}, \bibinfo {author}
  {\bibfnamefont {K.~W.}\ \bibnamefont {Madison}}, \bibinfo {author}
  {\bibfnamefont {S.}~\bibnamefont {Malbrunot-Ettenauer}}, \bibinfo {author}
  {\bibfnamefont {T.}~\bibnamefont {Miyagi}}, \bibinfo {author} {\bibfnamefont
  {I.~D.}\ \bibnamefont {Moore}}, \bibinfo {author} {\bibfnamefont
  {S.}~\bibnamefont {Moroch}}, \bibinfo {author} {\bibfnamefont
  {P.}~\bibnamefont {Navratil}}, \bibinfo {author} {\bibfnamefont
  {W.}~\bibnamefont {Nazarewicz}}, \bibinfo {author} {\bibfnamefont
  {G.}~\bibnamefont {Neyens}}, \bibinfo {author} {\bibfnamefont {E.~B.}\
  \bibnamefont {Norrgard}}, \bibinfo {author} {\bibfnamefont {N.}~\bibnamefont
  {Nusgart}}, \bibinfo {author} {\bibfnamefont {L.~F.}\ \bibnamefont
  {Pašteka}}, \bibinfo {author} {\bibfnamefont {A.~N.}\ \bibnamefont
  {Petrov}}, \bibinfo {author} {\bibfnamefont {W.~R.}\ \bibnamefont {Plaß}},
  \bibinfo {author} {\bibfnamefont {R.~A.}\ \bibnamefont {Ready}}, \bibinfo
  {author} {\bibfnamefont {M.~P.}\ \bibnamefont {Reiter}}, \bibinfo {author}
  {\bibfnamefont {M.}~\bibnamefont {Reponen}}, \bibinfo {author} {\bibfnamefont
  {S.}~\bibnamefont {Rothe}}, \bibinfo {author} {\bibfnamefont {M.~S.}\
  \bibnamefont {Safronova}}, \bibinfo {author} {\bibfnamefont {C.}~\bibnamefont
  {Scheidenerger}}, \bibinfo {author} {\bibfnamefont {A.}~\bibnamefont
  {Shindler}}, \bibinfo {author} {\bibfnamefont {J.~T.}\ \bibnamefont {Singh}},
  \bibinfo {author} {\bibfnamefont {L.~V.}\ \bibnamefont {Skripnikov}},
  \bibinfo {author} {\bibfnamefont {A.~V.}\ \bibnamefont {Titov}}, \bibinfo
  {author} {\bibfnamefont {S.-M.}\ \bibnamefont {Udrescu}}, \bibinfo {author}
  {\bibfnamefont {S.~G.}\ \bibnamefont {Wilkins}},\ and\ \bibinfo {author}
  {\bibfnamefont {X.}~\bibnamefont {Yang}},\ }\href
  {https://doi.org/10.1088/1361-6633/ad1e39} {\bibfield  {journal} {\bibinfo
  {journal} {Rep. Prog. Phys.}\ }\textbf {\bibinfo {volume} {87}},\ \bibinfo
  {pages} {084301} (\bibinfo {year} {2024})}\BibitemShut {NoStop}%
\bibitem [{\citenamefont {Khriplovich}(1991)}]{Khriplovich1991}%
  \BibitemOpen
  \bibfield  {author} {\bibinfo {author} {\bibfnamefont {I.~B.}\ \bibnamefont
  {Khriplovich}},\ }\href@noop {} {\emph {\bibinfo {title} {Parity
  Nonconservation in Atomic Phenomena}}}\ (\bibinfo  {publisher} {Gordon and
  Breach},\ \bibinfo {address} {Philadelphia},\ \bibinfo {year}
  {1991})\BibitemShut {NoStop}%
\bibitem [{\citenamefont {Khriplovich}\ and\ \citenamefont
  {Lamoureaux}(1997)}]{Khriplovich1997}%
  \BibitemOpen
  \bibfield  {author} {\bibinfo {author} {\bibfnamefont {I.~B.}\ \bibnamefont
  {Khriplovich}}\ and\ \bibinfo {author} {\bibfnamefont {S.~K.}\ \bibnamefont
  {Lamoureaux}},\ }\href@noop {} {\emph {\bibinfo {title} {CP Violation Without
  Strangenessa}}}\ (\bibinfo  {publisher} {Springer},\ \bibinfo {address}
  {Berlin},\ \bibinfo {year} {1997})\BibitemShut {NoStop}%
\bibitem [{\citenamefont {Dzuba}\ \emph {et~al.}(1999)\citenamefont {Dzuba},
  \citenamefont {Flambaum},\ and\ \citenamefont {Webb}}]{Dzuba1999}%
  \BibitemOpen
  \bibfield  {author} {\bibinfo {author} {\bibfnamefont {V.~A.}\ \bibnamefont
  {Dzuba}}, \bibinfo {author} {\bibfnamefont {V.~V.}\ \bibnamefont
  {Flambaum}},\ and\ \bibinfo {author} {\bibfnamefont {J.~K.}\ \bibnamefont
  {Webb}},\ }\href {https://doi.org/10.1103/PhysRevLett.82.888} {\bibfield
  {journal} {\bibinfo  {journal} {Phys. Rev. Lett.}\ }\textbf {\bibinfo
  {volume} {82}},\ \bibinfo {pages} {888} (\bibinfo {year} {1999})}\BibitemShut
  {NoStop}%
\bibitem [{\citenamefont {Berengut}\ \emph {et~al.}(2018)\citenamefont
  {Berengut}, \citenamefont {Budker}, \citenamefont {Delaunay}, \citenamefont
  {Flambaum}, \citenamefont {Frugiuele}, \citenamefont {Fuchs}, \citenamefont
  {Grojean}, \citenamefont {Harnik}, \citenamefont {Ozeri}, \citenamefont
  {Perez},\ and\ \citenamefont {Soreq}}]{Berengut2018}%
  \BibitemOpen
  \bibfield  {author} {\bibinfo {author} {\bibfnamefont {J.~C.}\ \bibnamefont
  {Berengut}}, \bibinfo {author} {\bibfnamefont {D.}~\bibnamefont {Budker}},
  \bibinfo {author} {\bibfnamefont {C.}~\bibnamefont {Delaunay}}, \bibinfo
  {author} {\bibfnamefont {V.~V.}\ \bibnamefont {Flambaum}}, \bibinfo {author}
  {\bibfnamefont {C.}~\bibnamefont {Frugiuele}}, \bibinfo {author}
  {\bibfnamefont {E.}~\bibnamefont {Fuchs}}, \bibinfo {author} {\bibfnamefont
  {C.}~\bibnamefont {Grojean}}, \bibinfo {author} {\bibfnamefont
  {R.}~\bibnamefont {Harnik}}, \bibinfo {author} {\bibfnamefont
  {R.}~\bibnamefont {Ozeri}}, \bibinfo {author} {\bibfnamefont
  {G.}~\bibnamefont {Perez}},\ and\ \bibinfo {author} {\bibfnamefont
  {Y.}~\bibnamefont {Soreq}},\ }\href
  {https://doi.org/10.1103/PhysRevLett.120.091801} {\bibfield  {journal}
  {\bibinfo  {journal} {Phys. Rev. Lett.}\ }\textbf {\bibinfo {volume} {120}},\
  \bibinfo {pages} {091801} (\bibinfo {year} {2018})}\BibitemShut {NoStop}%
\bibitem [{\citenamefont {Counts}\ \emph {et~al.}(2020)\citenamefont {Counts},
  \citenamefont {Hur}, \citenamefont {Aude~Craik}, \citenamefont {Jeon},
  \citenamefont {Leung}, \citenamefont {Berengut}, \citenamefont {Geddes},
  \citenamefont {Kawasaki}, \citenamefont {Jhe},\ and\ \citenamefont
  {Vuleti\ifmmode~\acute{c}\else \'{c}\fi{}}}]{Counts2020}%
  \BibitemOpen
  \bibfield  {author} {\bibinfo {author} {\bibfnamefont {I.}~\bibnamefont
  {Counts}}, \bibinfo {author} {\bibfnamefont {J.}~\bibnamefont {Hur}},
  \bibinfo {author} {\bibfnamefont {D.~P.~L.}\ \bibnamefont {Aude~Craik}},
  \bibinfo {author} {\bibfnamefont {H.}~\bibnamefont {Jeon}}, \bibinfo {author}
  {\bibfnamefont {C.}~\bibnamefont {Leung}}, \bibinfo {author} {\bibfnamefont
  {J.~C.}\ \bibnamefont {Berengut}}, \bibinfo {author} {\bibfnamefont
  {A.}~\bibnamefont {Geddes}}, \bibinfo {author} {\bibfnamefont
  {A.}~\bibnamefont {Kawasaki}}, \bibinfo {author} {\bibfnamefont
  {W.}~\bibnamefont {Jhe}},\ and\ \bibinfo {author} {\bibfnamefont
  {V.}~\bibnamefont {Vuleti\ifmmode~\acute{c}\else \'{c}\fi{}}},\ }\href
  {https://doi.org/10.1103/PhysRevLett.125.123002} {\bibfield  {journal}
  {\bibinfo  {journal} {Phys. Rev. Lett.}\ }\textbf {\bibinfo {volume} {125}},\
  \bibinfo {pages} {123002} (\bibinfo {year} {2020})}\BibitemShut {NoStop}%
\bibitem [{\citenamefont {Allehabi}\ \emph {et~al.}(2021)\citenamefont
  {Allehabi}, \citenamefont {Dzuba}, \citenamefont {Flambaum},\ and\
  \citenamefont {Afanasjev}}]{Allehabi2021}%
  \BibitemOpen
  \bibfield  {author} {\bibinfo {author} {\bibfnamefont {S.~O.}\ \bibnamefont
  {Allehabi}}, \bibinfo {author} {\bibfnamefont {V.~A.}\ \bibnamefont {Dzuba}},
  \bibinfo {author} {\bibfnamefont {V.~V.}\ \bibnamefont {Flambaum}},\ and\
  \bibinfo {author} {\bibfnamefont {A.~V.}\ \bibnamefont {Afanasjev}},\ }\href
  {https://doi.org/10.1103/PhysRevA.103.L030801} {\bibfield  {journal}
  {\bibinfo  {journal} {Phys. Rev. A}\ }\textbf {\bibinfo {volume} {103}},\
  \bibinfo {pages} {L030801} (\bibinfo {year} {2021})}\BibitemShut {NoStop}%
\bibitem [{\citenamefont {Flambaum}(1994)}]{Flambaum1994}%
  \BibitemOpen
  \bibfield  {author} {\bibinfo {author} {\bibfnamefont {V.~V.}\ \bibnamefont
  {Flambaum}},\ }\href
  {https://doi.org/https://doi.org/10.1016/0370-2693(94)90646-7} {\bibfield
  {journal} {\bibinfo  {journal} {Phys. Lett. B}\ }\textbf {\bibinfo {volume}
  {320}},\ \bibinfo {pages} {211} (\bibinfo {year} {1994})}\BibitemShut
  {NoStop}%
\bibitem [{\citenamefont {Auerbach}\ \emph {et~al.}(1996)\citenamefont
  {Auerbach}, \citenamefont {Flambaum},\ and\ \citenamefont
  {Spevak}}]{Auerbach1996}%
  \BibitemOpen
  \bibfield  {author} {\bibinfo {author} {\bibfnamefont {N.}~\bibnamefont
  {Auerbach}}, \bibinfo {author} {\bibfnamefont {V.~V.}\ \bibnamefont
  {Flambaum}},\ and\ \bibinfo {author} {\bibfnamefont {V.}~\bibnamefont
  {Spevak}},\ }\href {https://doi.org/10.1103/PhysRevLett.76.4316} {\bibfield
  {journal} {\bibinfo  {journal} {Phys. Rev. Lett.}\ }\textbf {\bibinfo
  {volume} {76}},\ \bibinfo {pages} {4316} (\bibinfo {year}
  {1996})}\BibitemShut {NoStop}%
\bibitem [{\citenamefont {Engel}\ \emph {et~al.}(2000)\citenamefont {Engel},
  \citenamefont {Friar},\ and\ \citenamefont {Hayes}}]{Engel2000}%
  \BibitemOpen
  \bibfield  {author} {\bibinfo {author} {\bibfnamefont {J.}~\bibnamefont
  {Engel}}, \bibinfo {author} {\bibfnamefont {J.~L.}\ \bibnamefont {Friar}},\
  and\ \bibinfo {author} {\bibfnamefont {A.~C.}\ \bibnamefont {Hayes}},\ }\href
  {https://doi.org/10.1103/PhysRevC.61.035502} {\bibfield  {journal} {\bibinfo
  {journal} {Phys. Rev. C}\ }\textbf {\bibinfo {volume} {61}},\ \bibinfo
  {pages} {035502} (\bibinfo {year} {2000})}\BibitemShut {NoStop}%
\bibitem [{\citenamefont {M\aa{}rtensson-Pendrill}(1995)}]{MP1995}%
  \BibitemOpen
  \bibfield  {author} {\bibinfo {author} {\bibfnamefont {A.-M.}\ \bibnamefont
  {M\aa{}rtensson-Pendrill}},\ }\href
  {https://doi.org/10.1103/PhysRevLett.74.2184} {\bibfield  {journal} {\bibinfo
   {journal} {Phys. Rev. Lett.}\ }\textbf {\bibinfo {volume} {74}},\ \bibinfo
  {pages} {2184} (\bibinfo {year} {1995})}\BibitemShut {NoStop}%
\bibitem [{\citenamefont {Ginges}\ \emph {et~al.}(2017)\citenamefont {Ginges},
  \citenamefont {Volotka},\ and\ \citenamefont {Fritzsche}}]{Ginges2017}%
  \BibitemOpen
  \bibfield  {author} {\bibinfo {author} {\bibfnamefont {J.~S.~M.}\
  \bibnamefont {Ginges}}, \bibinfo {author} {\bibfnamefont {A.~V.}\
  \bibnamefont {Volotka}},\ and\ \bibinfo {author} {\bibfnamefont
  {S.}~\bibnamefont {Fritzsche}},\ }\href
  {https://doi.org/10.1103/PhysRevA.96.062502} {\bibfield  {journal} {\bibinfo
  {journal} {Phys. Rev. A}\ }\textbf {\bibinfo {volume} {96}},\ \bibinfo
  {pages} {062502} (\bibinfo {year} {2017})}\BibitemShut {NoStop}%
\bibitem [{\citenamefont {Konovalova}\ \emph {et~al.}(2018)\citenamefont
  {Konovalova}, \citenamefont {Demidov}, \citenamefont {Kozlov},\ and\
  \citenamefont {Barzakh}}]{Konovalova2018}%
  \BibitemOpen
  \bibfield  {author} {\bibinfo {author} {\bibfnamefont {E.~A.}\ \bibnamefont
  {Konovalova}}, \bibinfo {author} {\bibfnamefont {Y.~A.}\ \bibnamefont
  {Demidov}}, \bibinfo {author} {\bibfnamefont {M.~G.}\ \bibnamefont
  {Kozlov}},\ and\ \bibinfo {author} {\bibfnamefont {A.~E.}\ \bibnamefont
  {Barzakh}},\ }\href {https://doi.org/10.3390/atoms6030039} {\bibfield
  {journal} {\bibinfo  {journal} {Atoms}\ }\textbf {\bibinfo {volume} {6}},\
  \bibinfo {pages} {39} (\bibinfo {year} {2018})}\BibitemShut {NoStop}%
\bibitem [{\citenamefont {Konovalova}\ \emph {et~al.}(2020)\citenamefont
  {Konovalova}, \citenamefont {Demidov},\ and\ \citenamefont
  {Kozlov}}]{Konovalova2020}%
  \BibitemOpen
  \bibfield  {author} {\bibinfo {author} {\bibfnamefont {E.~A.}\ \bibnamefont
  {Konovalova}}, \bibinfo {author} {\bibfnamefont {Y.~A.}\ \bibnamefont
  {Demidov}},\ and\ \bibinfo {author} {\bibfnamefont {M.~G.}\ \bibnamefont
  {Kozlov}},\ }\href {https://doi.org/10.1134/S0030400X20100148} {\bibfield
  {journal} {\bibinfo  {journal} {Opt. Spectrosc.}\ }\textbf {\bibinfo {volume}
  {128}},\ \bibinfo {pages} {1530} (\bibinfo {year} {2020})}\BibitemShut
  {NoStop}%
\bibitem [{\citenamefont {Roberts}\ and\ \citenamefont
  {Ginges}(2020)}]{RobertsFr2020}%
  \BibitemOpen
  \bibfield  {author} {\bibinfo {author} {\bibfnamefont {B.~M.}\ \bibnamefont
  {Roberts}}\ and\ \bibinfo {author} {\bibfnamefont {J.~S.~M.}\ \bibnamefont
  {Ginges}},\ }\href {https://doi.org/10.1103/PhysRevLett.125.063002}
  {\bibfield  {journal} {\bibinfo  {journal} {Phys. Rev. Lett.}\ }\textbf
  {\bibinfo {volume} {125}},\ \bibinfo {pages} {063002} (\bibinfo {year}
  {2020})}\BibitemShut {NoStop}%
\bibitem [{\citenamefont {Barzakh}\ \emph {et~al.}(2020)\citenamefont
  {Barzakh}, \citenamefont {Atanasov}, \citenamefont {Andreyev}, \citenamefont
  {Al~Monthery}, \citenamefont {Althubiti}, \citenamefont {Andel},
  \citenamefont {Antalic}, \citenamefont {Blaum}, \citenamefont {Cocolios},
  \citenamefont {Cubiss}, \citenamefont {Van~Duppen}, \citenamefont {Goodacre},
  \citenamefont {de~Roubin}, \citenamefont {Demidov}, \citenamefont
  {Farooq-Smith}, \citenamefont {Fedorov}, \citenamefont {Fedosseev},
  \citenamefont {Fink}, \citenamefont {Gaffney}, \citenamefont {Ghys},
  \citenamefont {Harding}, \citenamefont {Joss}, \citenamefont {Herfurth},
  \citenamefont {Huyse}, \citenamefont {Imai}, \citenamefont {Kozlov},
  \citenamefont {Kreim}, \citenamefont {Lunney}, \citenamefont {Lynch},
  \citenamefont {Manea}, \citenamefont {Marsh}, \citenamefont
  {Martinez~Palenzuela}, \citenamefont {Molkanov}, \citenamefont {Neidherr},
  \citenamefont {Page}, \citenamefont {Rosenbusch}, \citenamefont {Rossel},
  \citenamefont {Rothe}, \citenamefont {Schweikhard}, \citenamefont
  {Seliverstov}, \citenamefont {Sels}, \citenamefont {Van~Beveren},
  \citenamefont {Verstraelen}, \citenamefont {Welker}, \citenamefont
  {Wienholtz}, \citenamefont {Wolf},\ and\ \citenamefont
  {Zuber}}]{Barzakh2020}%
  \BibitemOpen
  \bibfield  {author} {\bibinfo {author} {\bibfnamefont {A.~E.}\ \bibnamefont
  {Barzakh}}, \bibinfo {author} {\bibfnamefont {D.}~\bibnamefont {Atanasov}},
  \bibinfo {author} {\bibfnamefont {A.~N.}\ \bibnamefont {Andreyev}}, \bibinfo
  {author} {\bibfnamefont {M.}~\bibnamefont {Al~Monthery}}, \bibinfo {author}
  {\bibfnamefont {N.~A.}\ \bibnamefont {Althubiti}}, \bibinfo {author}
  {\bibfnamefont {B.}~\bibnamefont {Andel}}, \bibinfo {author} {\bibfnamefont
  {S.}~\bibnamefont {Antalic}}, \bibinfo {author} {\bibfnamefont
  {K.}~\bibnamefont {Blaum}}, \bibinfo {author} {\bibfnamefont {T.~E.}\
  \bibnamefont {Cocolios}}, \bibinfo {author} {\bibfnamefont {J.~G.}\
  \bibnamefont {Cubiss}}, \bibinfo {author} {\bibfnamefont {P.}~\bibnamefont
  {Van~Duppen}}, \bibinfo {author} {\bibfnamefont {T.~D.}\ \bibnamefont
  {Goodacre}}, \bibinfo {author} {\bibfnamefont {A.}~\bibnamefont {de~Roubin}},
  \bibinfo {author} {\bibfnamefont {Y.~A.}\ \bibnamefont {Demidov}}, \bibinfo
  {author} {\bibfnamefont {G.~J.}\ \bibnamefont {Farooq-Smith}}, \bibinfo
  {author} {\bibfnamefont {D.~V.}\ \bibnamefont {Fedorov}}, \bibinfo {author}
  {\bibfnamefont {V.~N.}\ \bibnamefont {Fedosseev}}, \bibinfo {author}
  {\bibfnamefont {D.~A.}\ \bibnamefont {Fink}}, \bibinfo {author}
  {\bibfnamefont {L.~P.}\ \bibnamefont {Gaffney}}, \bibinfo {author}
  {\bibfnamefont {L.}~\bibnamefont {Ghys}}, \bibinfo {author} {\bibfnamefont
  {R.~D.}\ \bibnamefont {Harding}}, \bibinfo {author} {\bibfnamefont {D.~T.}\
  \bibnamefont {Joss}}, \bibinfo {author} {\bibfnamefont {F.}~\bibnamefont
  {Herfurth}}, \bibinfo {author} {\bibfnamefont {M.}~\bibnamefont {Huyse}},
  \bibinfo {author} {\bibfnamefont {N.}~\bibnamefont {Imai}}, \bibinfo {author}
  {\bibfnamefont {M.~G.}\ \bibnamefont {Kozlov}}, \bibinfo {author}
  {\bibfnamefont {S.}~\bibnamefont {Kreim}}, \bibinfo {author} {\bibfnamefont
  {D.}~\bibnamefont {Lunney}}, \bibinfo {author} {\bibfnamefont {K.~M.}\
  \bibnamefont {Lynch}}, \bibinfo {author} {\bibfnamefont {V.}~\bibnamefont
  {Manea}}, \bibinfo {author} {\bibfnamefont {B.~A.}\ \bibnamefont {Marsh}},
  \bibinfo {author} {\bibfnamefont {Y.}~\bibnamefont {Martinez~Palenzuela}},
  \bibinfo {author} {\bibfnamefont {P.~L.}\ \bibnamefont {Molkanov}}, \bibinfo
  {author} {\bibfnamefont {D.}~\bibnamefont {Neidherr}}, \bibinfo {author}
  {\bibfnamefont {R.~D.}\ \bibnamefont {Page}}, \bibinfo {author}
  {\bibfnamefont {M.}~\bibnamefont {Rosenbusch}}, \bibinfo {author}
  {\bibfnamefont {R.~E.}\ \bibnamefont {Rossel}}, \bibinfo {author}
  {\bibfnamefont {S.}~\bibnamefont {Rothe}}, \bibinfo {author} {\bibfnamefont
  {L.}~\bibnamefont {Schweikhard}}, \bibinfo {author} {\bibfnamefont {M.~D.}\
  \bibnamefont {Seliverstov}}, \bibinfo {author} {\bibfnamefont
  {S.}~\bibnamefont {Sels}}, \bibinfo {author} {\bibfnamefont {C.}~\bibnamefont
  {Van~Beveren}}, \bibinfo {author} {\bibfnamefont {E.}~\bibnamefont
  {Verstraelen}}, \bibinfo {author} {\bibfnamefont {A.}~\bibnamefont {Welker}},
  \bibinfo {author} {\bibfnamefont {F.}~\bibnamefont {Wienholtz}}, \bibinfo
  {author} {\bibfnamefont {R.~N.}\ \bibnamefont {Wolf}},\ and\ \bibinfo
  {author} {\bibfnamefont {K.}~\bibnamefont {Zuber}},\ }\href
  {https://doi.org/10.1103/PhysRevC.101.034308} {\bibfield  {journal} {\bibinfo
   {journal} {Phys. Rev. C}\ }\textbf {\bibinfo {volume} {101}},\ \bibinfo
  {pages} {034308} (\bibinfo {year} {2020})}\BibitemShut {NoStop}%
\bibitem [{\citenamefont {Demidov}\ \emph {et~al.}(2021)\citenamefont
  {Demidov}, \citenamefont {Konovalova}, \citenamefont {Imanbaeva},
  \citenamefont {Kozlov},\ and\ \citenamefont {Barzakh}}]{Demidov2021}%
  \BibitemOpen
  \bibfield  {author} {\bibinfo {author} {\bibfnamefont {Y.~A.}\ \bibnamefont
  {Demidov}}, \bibinfo {author} {\bibfnamefont {E.~A.}\ \bibnamefont
  {Konovalova}}, \bibinfo {author} {\bibfnamefont {R.~T.}\ \bibnamefont
  {Imanbaeva}}, \bibinfo {author} {\bibfnamefont {M.~G.}\ \bibnamefont
  {Kozlov}},\ and\ \bibinfo {author} {\bibfnamefont {A.~E.}\ \bibnamefont
  {Barzakh}},\ }\href {https://doi.org/10.1103/PhysRevA.103.032824} {\bibfield
  {journal} {\bibinfo  {journal} {Phys. Rev. A}\ }\textbf {\bibinfo {volume}
  {103}},\ \bibinfo {pages} {032824} (\bibinfo {year} {2021})}\BibitemShut
  {NoStop}%
\bibitem [{\citenamefont {Roberts}\ and\ \citenamefont
  {Ginges}(2021)}]{Roberts2021}%
  \BibitemOpen
  \bibfield  {author} {\bibinfo {author} {\bibfnamefont {B.~M.}\ \bibnamefont
  {Roberts}}\ and\ \bibinfo {author} {\bibfnamefont {J.~S.~M.}\ \bibnamefont
  {Ginges}},\ }\href {https://doi.org/10.1103/PhysRevA.104.022823} {\bibfield
  {journal} {\bibinfo  {journal} {Phys. Rev. A}\ }\textbf {\bibinfo {volume}
  {104}},\ \bibinfo {pages} {022823} (\bibinfo {year} {2021})}\BibitemShut
  {NoStop}%
\bibitem [{\citenamefont {Skripnikov}(2020)}]{Skripnikov2020}%
  \BibitemOpen
  \bibfield  {author} {\bibinfo {author} {\bibfnamefont {L.~V.}\ \bibnamefont
  {Skripnikov}},\ }\href {https://doi.org/10.1063/5.0024103} {\bibfield
  {journal} {\bibinfo  {journal} {J. Chem. Phys.}\ }\textbf {\bibinfo {volume}
  {153}},\ \bibinfo {pages} {114114} (\bibinfo {year} {2020})}\BibitemShut
  {NoStop}%
\bibitem [{\citenamefont {Roberts}\ \emph {et~al.}(2022)\citenamefont
  {Roberts}, \citenamefont {Ranclaud},\ and\ \citenamefont
  {Ginges}}]{Roberts2021scr}%
  \BibitemOpen
  \bibfield  {author} {\bibinfo {author} {\bibfnamefont {B.~M.}\ \bibnamefont
  {Roberts}}, \bibinfo {author} {\bibfnamefont {P.~G.}\ \bibnamefont
  {Ranclaud}},\ and\ \bibinfo {author} {\bibfnamefont {J.~S.~M.}\ \bibnamefont
  {Ginges}},\ }\href {https://doi.org/10.1103/PhysRevA.105.052802} {\bibfield
  {journal} {\bibinfo  {journal} {Phys. Rev. A}\ }\textbf {\bibinfo {volume}
  {105}},\ \bibinfo {pages} {052802} (\bibinfo {year} {2022})}\BibitemShut
  {NoStop}%
\bibitem [{\citenamefont {Prosnyak}\ and\ \citenamefont
  {Skripnikov}(2021)}]{Prosnyak2021}%
  \BibitemOpen
  \bibfield  {author} {\bibinfo {author} {\bibfnamefont {S.~D.}\ \bibnamefont
  {Prosnyak}}\ and\ \bibinfo {author} {\bibfnamefont {L.~V.}\ \bibnamefont
  {Skripnikov}},\ }\href {https://doi.org/10.1103/PhysRevC.103.034314}
  {\bibfield  {journal} {\bibinfo  {journal} {Phys. Rev. C}\ }\textbf {\bibinfo
  {volume} {103}},\ \bibinfo {pages} {034314} (\bibinfo {year}
  {2021})}\BibitemShut {NoStop}%
\bibitem [{\citenamefont {Skripnikov}\ and\ \citenamefont
  {Prosnyak}(2022)}]{Skripnikov2022}%
  \BibitemOpen
  \bibfield  {author} {\bibinfo {author} {\bibfnamefont {L.~V.}\ \bibnamefont
  {Skripnikov}}\ and\ \bibinfo {author} {\bibfnamefont {S.~D.}\ \bibnamefont
  {Prosnyak}},\ }\href {https://doi.org/10.1103/PhysRevC.106.054303} {\bibfield
   {journal} {\bibinfo  {journal} {Phys. Rev. C}\ }\textbf {\bibinfo {volume}
  {106}},\ \bibinfo {pages} {054303} (\bibinfo {year} {2022})}\BibitemShut
  {NoStop}%
\bibitem [{\citenamefont {Sanamyan}\ \emph {et~al.}(2023)\citenamefont
  {Sanamyan}, \citenamefont {Roberts},\ and\ \citenamefont
  {Ginges}}]{Sanamyan2023}%
  \BibitemOpen
  \bibfield  {author} {\bibinfo {author} {\bibfnamefont {G.}~\bibnamefont
  {Sanamyan}}, \bibinfo {author} {\bibfnamefont {B.~M.}\ \bibnamefont
  {Roberts}},\ and\ \bibinfo {author} {\bibfnamefont {J.~S.~M.}\ \bibnamefont
  {Ginges}},\ }\href {https://doi.org/10.1103/physrevlett.130.053001}
  {\bibfield  {journal} {\bibinfo  {journal} {Phys. Rev. Lett.}\ }\textbf
  {\bibinfo {volume} {130}},\ \bibinfo {pages} {053001} (\bibinfo {year}
  {2023})}\BibitemShut {NoStop}%
\bibitem [{\citenamefont {Lechner}\ \emph {et~al.}(2023)\citenamefont
  {Lechner}, \citenamefont {Miyagi}, \citenamefont {Xu}, \citenamefont
  {Bissell}, \citenamefont {Blaum}, \citenamefont {Cheal}, \citenamefont
  {Devlin}, \citenamefont {{Garcia Ruiz}}, \citenamefont {Ginges},
  \citenamefont {Heylen}, \citenamefont {Holt}, \citenamefont {Imgram},
  \citenamefont {Kanellakopoulos}, \citenamefont {Koszor{\'u}s}, \citenamefont
  {Malbrunot-Ettenauer}, \citenamefont {Neugart}, \citenamefont {Neyens},
  \citenamefont {N{\"o}rtersh{\"a}user}, \citenamefont {Plattner},
  \citenamefont {Rodr{\'i}guez}, \citenamefont {Sanamyan}, \citenamefont
  {Stroberg}, \citenamefont {Utsuno}, \citenamefont {Yang},\ and\ \citenamefont
  {Yordanov}}]{Lechner2023}%
  \BibitemOpen
  \bibfield  {author} {\bibinfo {author} {\bibfnamefont {S.}~\bibnamefont
  {Lechner}}, \bibinfo {author} {\bibfnamefont {T.}~\bibnamefont {Miyagi}},
  \bibinfo {author} {\bibfnamefont {Z.~Y.}\ \bibnamefont {Xu}}, \bibinfo
  {author} {\bibfnamefont {M.~L.}\ \bibnamefont {Bissell}}, \bibinfo {author}
  {\bibfnamefont {K.}~\bibnamefont {Blaum}}, \bibinfo {author} {\bibfnamefont
  {B.}~\bibnamefont {Cheal}}, \bibinfo {author} {\bibfnamefont {C.~S.}\
  \bibnamefont {Devlin}}, \bibinfo {author} {\bibfnamefont {R.~F.}\
  \bibnamefont {{Garcia Ruiz}}}, \bibinfo {author} {\bibfnamefont {J.~S.~M.}\
  \bibnamefont {Ginges}}, \bibinfo {author} {\bibfnamefont {H.}~\bibnamefont
  {Heylen}}, \bibinfo {author} {\bibfnamefont {J.~D.}\ \bibnamefont {Holt}},
  \bibinfo {author} {\bibfnamefont {P.}~\bibnamefont {Imgram}}, \bibinfo
  {author} {\bibfnamefont {A.}~\bibnamefont {Kanellakopoulos}}, \bibinfo
  {author} {\bibfnamefont {{\'A}.}~\bibnamefont {Koszor{\'u}s}}, \bibinfo
  {author} {\bibfnamefont {S.}~\bibnamefont {Malbrunot-Ettenauer}}, \bibinfo
  {author} {\bibfnamefont {R.}~\bibnamefont {Neugart}}, \bibinfo {author}
  {\bibfnamefont {G.}~\bibnamefont {Neyens}}, \bibinfo {author} {\bibfnamefont
  {W.}~\bibnamefont {N{\"o}rtersh{\"a}user}}, \bibinfo {author} {\bibfnamefont
  {P.}~\bibnamefont {Plattner}}, \bibinfo {author} {\bibfnamefont {L.~V.}\
  \bibnamefont {Rodr{\'i}guez}}, \bibinfo {author} {\bibfnamefont
  {G.}~\bibnamefont {Sanamyan}}, \bibinfo {author} {\bibfnamefont {S.~R.}\
  \bibnamefont {Stroberg}}, \bibinfo {author} {\bibfnamefont {Y.}~\bibnamefont
  {Utsuno}}, \bibinfo {author} {\bibfnamefont {X.~F.}\ \bibnamefont {Yang}},\
  and\ \bibinfo {author} {\bibfnamefont {D.~T.}\ \bibnamefont {Yordanov}},\
  }\href {https://doi.org/https://doi.org/10.1016/j.physletb.2023.138278}
  {\bibfield  {journal} {\bibinfo  {journal} {Phys. Lett. B}\ }\textbf
  {\bibinfo {volume} {847}},\ \bibinfo {pages} {138278} (\bibinfo {year}
  {2023})}\BibitemShut {NoStop}%
\bibitem [{\citenamefont {Wilkins}\ \emph {et~al.}(2023)\citenamefont
  {Wilkins}, \citenamefont {Udrescu}, \citenamefont {Athanasakis-Kaklamanakis},
  \citenamefont {Ruiz}, \citenamefont {Au}, \citenamefont
  {Belo{\v{s}}evi{\'{c}}}, \citenamefont {Berger}, \citenamefont {Bissell},
  \citenamefont {Breier}, \citenamefont {Brinson}, \citenamefont {Chrysalidis},
  \citenamefont {Cocolios}, \citenamefont {de~Groote}, \citenamefont {Dorne},
  \citenamefont {Flanagan}, \citenamefont {Franchoo}, \citenamefont {Gaul},
  \citenamefont {Geldhof}, \citenamefont {Giesen}, \citenamefont {Hanstorp},
  \citenamefont {Heinke}, \citenamefont {Isaev}, \citenamefont {Koszor{\'u}s},
  \citenamefont {Kujanp{\"a}{\"a}}, \citenamefont {Lalanne}, \citenamefont
  {Neyens}, \citenamefont {Nichols}, \citenamefont {Perrett}, \citenamefont
  {Reilly}, \citenamefont {Skripnikov}, \citenamefont {Rothe}, \citenamefont
  {van~den Borne}, \citenamefont {Wang}, \citenamefont {Wessolek},
  \citenamefont {Yang},\ and\ \citenamefont {Z{\"u}lch}}]{Wilkins2023}%
  \BibitemOpen
  \bibfield  {author} {\bibinfo {author} {\bibfnamefont {S.~G.}\ \bibnamefont
  {Wilkins}}, \bibinfo {author} {\bibfnamefont {S.~M.}\ \bibnamefont
  {Udrescu}}, \bibinfo {author} {\bibfnamefont {M.}~\bibnamefont
  {Athanasakis-Kaklamanakis}}, \bibinfo {author} {\bibfnamefont {R.~F.~G.}\
  \bibnamefont {Ruiz}}, \bibinfo {author} {\bibfnamefont {M.}~\bibnamefont
  {Au}}, \bibinfo {author} {\bibfnamefont {I.}~\bibnamefont
  {Belo{\v{s}}evi{\'{c}}}}, \bibinfo {author} {\bibfnamefont {R.}~\bibnamefont
  {Berger}}, \bibinfo {author} {\bibfnamefont {M.~L.}\ \bibnamefont {Bissell}},
  \bibinfo {author} {\bibfnamefont {A.~A.}\ \bibnamefont {Breier}}, \bibinfo
  {author} {\bibfnamefont {A.~J.}\ \bibnamefont {Brinson}}, \bibinfo {author}
  {\bibfnamefont {K.}~\bibnamefont {Chrysalidis}}, \bibinfo {author}
  {\bibfnamefont {T.~E.}\ \bibnamefont {Cocolios}}, \bibinfo {author}
  {\bibfnamefont {R.~P.}\ \bibnamefont {de~Groote}}, \bibinfo {author}
  {\bibfnamefont {A.}~\bibnamefont {Dorne}}, \bibinfo {author} {\bibfnamefont
  {K.~T.}\ \bibnamefont {Flanagan}}, \bibinfo {author} {\bibfnamefont
  {S.}~\bibnamefont {Franchoo}}, \bibinfo {author} {\bibfnamefont
  {K.}~\bibnamefont {Gaul}}, \bibinfo {author} {\bibfnamefont {S.}~\bibnamefont
  {Geldhof}}, \bibinfo {author} {\bibfnamefont {T.~F.}\ \bibnamefont {Giesen}},
  \bibinfo {author} {\bibfnamefont {D.}~\bibnamefont {Hanstorp}}, \bibinfo
  {author} {\bibfnamefont {R.}~\bibnamefont {Heinke}}, \bibinfo {author}
  {\bibfnamefont {T.}~\bibnamefont {Isaev}}, \bibinfo {author} {\bibfnamefont
  {{\'A}.}~\bibnamefont {Koszor{\'u}s}}, \bibinfo {author} {\bibfnamefont
  {S.}~\bibnamefont {Kujanp{\"a}{\"a}}}, \bibinfo {author} {\bibfnamefont
  {L.}~\bibnamefont {Lalanne}}, \bibinfo {author} {\bibfnamefont
  {G.}~\bibnamefont {Neyens}}, \bibinfo {author} {\bibfnamefont
  {M.}~\bibnamefont {Nichols}}, \bibinfo {author} {\bibfnamefont {H.~A.}\
  \bibnamefont {Perrett}}, \bibinfo {author} {\bibfnamefont {J.~R.}\
  \bibnamefont {Reilly}}, \bibinfo {author} {\bibfnamefont {L.~V.}\
  \bibnamefont {Skripnikov}}, \bibinfo {author} {\bibfnamefont
  {S.}~\bibnamefont {Rothe}}, \bibinfo {author} {\bibfnamefont
  {B.}~\bibnamefont {van~den Borne}}, \bibinfo {author} {\bibfnamefont
  {Q.}~\bibnamefont {Wang}}, \bibinfo {author} {\bibfnamefont {J.}~\bibnamefont
  {Wessolek}}, \bibinfo {author} {\bibfnamefont {X.~F.}\ \bibnamefont {Yang}},\
  and\ \bibinfo {author} {\bibfnamefont {C.}~\bibnamefont {Z{\"u}lch}},\ }\href
  {https://arxiv.org/abs/2311.04121} {} (\bibinfo {year} {2023}),\ \Eprint
  {https://arxiv.org/abs/2311.04121} {arXiv:2311.04121} \BibitemShut {NoStop}%
\bibitem [{\citenamefont {Vandeleur}\ \emph {et~al.}(2024)\citenamefont
  {Vandeleur}, \citenamefont {Sanamyan}, \citenamefont {Smits}, \citenamefont
  {Valuev}, \citenamefont {Oreshkina},\ and\ \citenamefont
  {Ginges}}]{Vandeleur2024}%
  \BibitemOpen
  \bibfield  {author} {\bibinfo {author} {\bibfnamefont {J.}~\bibnamefont
  {Vandeleur}}, \bibinfo {author} {\bibfnamefont {G.}~\bibnamefont {Sanamyan}},
  \bibinfo {author} {\bibfnamefont {O.~R.}\ \bibnamefont {Smits}}, \bibinfo
  {author} {\bibfnamefont {I.~A.}\ \bibnamefont {Valuev}}, \bibinfo {author}
  {\bibfnamefont {N.~S.}\ \bibnamefont {Oreshkina}},\ and\ \bibinfo {author}
  {\bibfnamefont {J.~S.~M.}\ \bibnamefont {Ginges}},\ }\href
  {https://arxiv.org/abs/2408.16516} {} (\bibinfo {year} {2024}),\ \Eprint
  {https://arxiv.org/abs/2408.16516} {arXiv:2408.16516} \BibitemShut {NoStop}%
\bibitem [{\citenamefont {Bohr}\ and\ \citenamefont
  {Weisskopf}(1950)}]{Bohr1950}%
  \BibitemOpen
  \bibfield  {author} {\bibinfo {author} {\bibfnamefont {A.}~\bibnamefont
  {Bohr}}\ and\ \bibinfo {author} {\bibfnamefont {V.~F.}\ \bibnamefont
  {Weisskopf}},\ }\href {https://doi.org/10.1103/PhysRev.77.94} {\bibfield
  {journal} {\bibinfo  {journal} {Phys. Rev.}\ }\textbf {\bibinfo {volume}
  {77}},\ \bibinfo {pages} {94} (\bibinfo {year} {1950})}\BibitemShut {NoStop}%
\bibitem [{\citenamefont {Bohr}(1951)}]{Bohr1951}%
  \BibitemOpen
  \bibfield  {author} {\bibinfo {author} {\bibfnamefont {A.}~\bibnamefont
  {Bohr}},\ }\href {https://doi.org/10.1103/PhysRev.81.331} {\bibfield
  {journal} {\bibinfo  {journal} {Phys. Rev.}\ }\textbf {\bibinfo {volume}
  {81}},\ \bibinfo {pages} {331} (\bibinfo {year} {1951})}\BibitemShut
  {NoStop}%
\bibitem [{\citenamefont {Fujita}\ and\ \citenamefont
  {Arima}(1975)}]{Fujita1975}%
  \BibitemOpen
  \bibfield  {author} {\bibinfo {author} {\bibfnamefont {T.}~\bibnamefont
  {Fujita}}\ and\ \bibinfo {author} {\bibfnamefont {A.}~\bibnamefont {Arima}},\
  }\href {https://doi.org/10.1016/0375-9474(75)90234-1} {\bibfield  {journal}
  {\bibinfo  {journal} {Nuc. Phys. A}\ }\textbf {\bibinfo {volume} {254}},\
  \bibinfo {pages} {513} (\bibinfo {year} {1975})}\BibitemShut {NoStop}%
\bibitem [{\citenamefont {B{\"{u}}ttgenbach}(1984)}]{Buttgenbach1984}%
  \BibitemOpen
  \bibfield  {author} {\bibinfo {author} {\bibfnamefont {S.}~\bibnamefont
  {B{\"{u}}ttgenbach}},\ }\href {https://doi.org/10.1007/BF02043319} {\bibfield
   {journal} {\bibinfo  {journal} {Hyperfine Interact.}\ }\textbf {\bibinfo
  {volume} {20}},\ \bibinfo {pages} {1} (\bibinfo {year} {1984})}\BibitemShut
  {NoStop}%
\bibitem [{\citenamefont {Shabaev}(1994)}]{Shabaev1994}%
  \BibitemOpen
  \bibfield  {author} {\bibinfo {author} {\bibfnamefont {V.~M.}\ \bibnamefont
  {Shabaev}},\ }\href {https://doi.org/10.1088/0953-4075/27/24/006} {\bibfield
  {journal} {\bibinfo  {journal} {J. Phys. B}\ }\textbf {\bibinfo {volume}
  {27}},\ \bibinfo {pages} {5825} (\bibinfo {year} {1994})}\BibitemShut
  {NoStop}%
\bibitem [{\citenamefont {Sen'kov}\ and\ \citenamefont
  {Dmitriev}(2002)}]{Senkov2002}%
  \BibitemOpen
  \bibfield  {author} {\bibinfo {author} {\bibfnamefont {R.~A.}\ \bibnamefont
  {Sen'kov}}\ and\ \bibinfo {author} {\bibfnamefont {V.~F.}\ \bibnamefont
  {Dmitriev}},\ }\href
  {https://doi.org/https://doi.org/10.1016/S0375-9474(02)00759-5} {\bibfield
  {journal} {\bibinfo  {journal} {Nucl. Phys. A}\ }\textbf {\bibinfo {volume}
  {706}},\ \bibinfo {pages} {351} (\bibinfo {year} {2002})}\BibitemShut
  {NoStop}%
\bibitem [{\citenamefont {Tomaselli}\ \emph {et~al.}(2002)\citenamefont
  {Tomaselli}, \citenamefont {K\"uhl}, \citenamefont {N\"ortersh\"auser},
  \citenamefont {Borneis}, \citenamefont {Dax}, \citenamefont {Marx},
  \citenamefont {Wang},\ and\ \citenamefont {Fritzsche}}]{Tomaselli2002}%
  \BibitemOpen
  \bibfield  {author} {\bibinfo {author} {\bibfnamefont {M.}~\bibnamefont
  {Tomaselli}}, \bibinfo {author} {\bibfnamefont {T.}~\bibnamefont {K\"uhl}},
  \bibinfo {author} {\bibfnamefont {W.}~\bibnamefont {N\"ortersh\"auser}},
  \bibinfo {author} {\bibfnamefont {S.}~\bibnamefont {Borneis}}, \bibinfo
  {author} {\bibfnamefont {A.}~\bibnamefont {Dax}}, \bibinfo {author}
  {\bibfnamefont {D.}~\bibnamefont {Marx}}, \bibinfo {author} {\bibfnamefont
  {H.}~\bibnamefont {Wang}},\ and\ \bibinfo {author} {\bibfnamefont
  {S.}~\bibnamefont {Fritzsche}},\ }\href
  {https://doi.org/10.1103/PhysRevA.65.022502} {\bibfield  {journal} {\bibinfo
  {journal} {Phys. Rev. A}\ }\textbf {\bibinfo {volume} {65}},\ \bibinfo
  {pages} {022502} (\bibinfo {year} {2002})}\BibitemShut {NoStop}%
\bibitem [{\citenamefont {Moskowitz}\ and\ \citenamefont
  {Lombardi}(1973)}]{Moskowitz1973}%
  \BibitemOpen
  \bibfield  {author} {\bibinfo {author} {\bibfnamefont {P.~A.}\ \bibnamefont
  {Moskowitz}}\ and\ \bibinfo {author} {\bibfnamefont {M.}~\bibnamefont
  {Lombardi}},\ }\href {https://doi.org/10.1016/0370-2693(73)90132-9}
  {\bibfield  {journal} {\bibinfo  {journal} {Phys. Lett. B}\ }\textbf
  {\bibinfo {volume} {46}},\ \bibinfo {pages} {334} (\bibinfo {year}
  {1973})}\BibitemShut {NoStop}%
\bibitem [{\citenamefont {Moskowitz}(1982)}]{Moskowitz1982}%
  \BibitemOpen
  \bibfield  {author} {\bibinfo {author} {\bibfnamefont {P.~A.}\ \bibnamefont
  {Moskowitz}},\ }\href {https://doi.org/10.1016/0370-2693(82)90595-0}
  {\bibfield  {journal} {\bibinfo  {journal} {Phys. Lett. B}\ }\textbf
  {\bibinfo {volume} {118}},\ \bibinfo {pages} {29} (\bibinfo {year}
  {1982})}\BibitemShut {NoStop}%
\bibitem [{\citenamefont {Fr{\"o}mmgen}\ \emph {et~al.}(2015)\citenamefont
  {Fr{\"o}mmgen}, \citenamefont {Balabanski}, \citenamefont {Bissell},
  \citenamefont {Biero{\'n}}, \citenamefont {Blaum}, \citenamefont {Cheal},
  \citenamefont {Flanagan}, \citenamefont {Fritzsche}, \citenamefont {Geppert},
  \citenamefont {Hammen}, \citenamefont {Kowalska}, \citenamefont {Kreim},
  \citenamefont {Krieger}, \citenamefont {Neugart}, \citenamefont {Neyens},
  \citenamefont {Rajabali}, \citenamefont {N{\"o}rtersh{\"a}user},
  \citenamefont {Papuga},\ and\ \citenamefont {Yordanov}}]{Frommgen2015}%
  \BibitemOpen
  \bibfield  {author} {\bibinfo {author} {\bibfnamefont {N.}~\bibnamefont
  {Fr{\"o}mmgen}}, \bibinfo {author} {\bibfnamefont {D.~L.}\ \bibnamefont
  {Balabanski}}, \bibinfo {author} {\bibfnamefont {M.~L.}\ \bibnamefont
  {Bissell}}, \bibinfo {author} {\bibfnamefont {J.}~\bibnamefont {Biero{\'n}}},
  \bibinfo {author} {\bibfnamefont {K.}~\bibnamefont {Blaum}}, \bibinfo
  {author} {\bibfnamefont {B.}~\bibnamefont {Cheal}}, \bibinfo {author}
  {\bibfnamefont {K.}~\bibnamefont {Flanagan}}, \bibinfo {author}
  {\bibfnamefont {S.}~\bibnamefont {Fritzsche}}, \bibinfo {author}
  {\bibfnamefont {C.}~\bibnamefont {Geppert}}, \bibinfo {author} {\bibfnamefont
  {M.}~\bibnamefont {Hammen}}, \bibinfo {author} {\bibfnamefont
  {M.}~\bibnamefont {Kowalska}}, \bibinfo {author} {\bibfnamefont
  {K.}~\bibnamefont {Kreim}}, \bibinfo {author} {\bibfnamefont
  {A.}~\bibnamefont {Krieger}}, \bibinfo {author} {\bibfnamefont
  {R.}~\bibnamefont {Neugart}}, \bibinfo {author} {\bibfnamefont
  {G.}~\bibnamefont {Neyens}}, \bibinfo {author} {\bibfnamefont {M.~M.}\
  \bibnamefont {Rajabali}}, \bibinfo {author} {\bibfnamefont {W.}~\bibnamefont
  {N{\"o}rtersh{\"a}user}}, \bibinfo {author} {\bibfnamefont {J.}~\bibnamefont
  {Papuga}},\ and\ \bibinfo {author} {\bibfnamefont {D.~T.}\ \bibnamefont
  {Yordanov}},\ }\href {https://doi.org/10.1140/epjd/e2015-60219-0} {\bibfield
  {journal} {\bibinfo  {journal} {Eur. Phys. J. D}\ }\textbf {\bibinfo {volume}
  {69}},\ \bibinfo {pages} {164} (\bibinfo {year} {2015})}\BibitemShut
  {NoStop}%
\bibitem [{\citenamefont {Persson}(2020)}]{Persson2020}%
  \BibitemOpen
  \bibfield  {author} {\bibinfo {author} {\bibfnamefont {J.~R.}\ \bibnamefont
  {Persson}},\ }\href {https://doi.org/10.3390/atoms8010005} {\bibfield
  {journal} {\bibinfo  {journal} {Atoms}\ }\textbf {\bibinfo {volume} {8}},\
  \bibinfo {pages} {5} (\bibinfo {year} {2020})}\BibitemShut {NoStop}%
\bibitem [{\citenamefont {Graner}\ \emph {et~al.}(2016)\citenamefont {Graner},
  \citenamefont {Chen}, \citenamefont {Lindahl},\ and\ \citenamefont
  {Heckel}}]{Graner2016}%
  \BibitemOpen
  \bibfield  {author} {\bibinfo {author} {\bibfnamefont {B.}~\bibnamefont
  {Graner}}, \bibinfo {author} {\bibfnamefont {Y.}~\bibnamefont {Chen}},
  \bibinfo {author} {\bibfnamefont {E.}~\bibnamefont {Lindahl}},\ and\ \bibinfo
  {author} {\bibfnamefont {B.}~\bibnamefont {Heckel}},\ }\href
  {https://doi.org/10.1103/physrevlett.116.161601} {\bibfield  {journal}
  {\bibinfo  {journal} {Phys. Rev. Lett.}\ }\textbf {\bibinfo {volume} {116}},\
  \bibinfo {pages} {161601} (\bibinfo {year} {2016})}\BibitemShut {NoStop}%
\bibitem [{\citenamefont {Engel}\ \emph {et~al.}(2013)\citenamefont {Engel},
  \citenamefont {Ramsey-Musolf},\ and\ \citenamefont {{van
  Kolck}}}]{Engel2013}%
  \BibitemOpen
  \bibfield  {author} {\bibinfo {author} {\bibfnamefont {J.}~\bibnamefont
  {Engel}}, \bibinfo {author} {\bibfnamefont {M.~J.}\ \bibnamefont
  {Ramsey-Musolf}},\ and\ \bibinfo {author} {\bibfnamefont {U.}~\bibnamefont
  {{van Kolck}}},\ }\href
  {https://doi.org/https://doi.org/10.1016/j.ppnp.2013.03.003} {\bibfield
  {journal} {\bibinfo  {journal} {Progress in Particle and Nuclear Physics}\
  }\textbf {\bibinfo {volume} {71}},\ \bibinfo {pages} {21} (\bibinfo {year}
  {2013})}\BibitemShut {NoStop}%
\bibitem [{\citenamefont {Berkeland}\ \emph {et~al.}(1998)\citenamefont
  {Berkeland}, \citenamefont {Miller}, \citenamefont {Bergquist}, \citenamefont
  {Itano},\ and\ \citenamefont {Wineland}}]{Berkeland1998}%
  \BibitemOpen
  \bibfield  {author} {\bibinfo {author} {\bibfnamefont {D.~J.}\ \bibnamefont
  {Berkeland}}, \bibinfo {author} {\bibfnamefont {J.~D.}\ \bibnamefont
  {Miller}}, \bibinfo {author} {\bibfnamefont {J.~C.}\ \bibnamefont
  {Bergquist}}, \bibinfo {author} {\bibfnamefont {W.~M.}\ \bibnamefont
  {Itano}},\ and\ \bibinfo {author} {\bibfnamefont {D.~J.}\ \bibnamefont
  {Wineland}},\ }\href {https://doi.org/10.1103/PhysRevLett.80.2089} {\bibfield
   {journal} {\bibinfo  {journal} {Phys. Rev. Lett.}\ }\textbf {\bibinfo
  {volume} {80}},\ \bibinfo {pages} {2089} (\bibinfo {year}
  {1998})}\BibitemShut {NoStop}%
\bibitem [{\citenamefont {Diddams}(2001)}]{Diddams2001}%
  \BibitemOpen
  \bibfield  {author} {\bibinfo {author} {\bibfnamefont {S.~A.}\ \bibnamefont
  {Diddams}},\ }\href {https://doi.org/10.1126/science.1061171} {\bibfield
  {journal} {\bibinfo  {journal} {Science}\ }\textbf {\bibinfo {volume}
  {293}},\ \bibinfo {pages} {825} (\bibinfo {year} {2001})}\BibitemShut
  {NoStop}%
\bibitem [{\citenamefont {Hachisu}\ \emph {et~al.}(2008)\citenamefont
  {Hachisu}, \citenamefont {Miyagishi}, \citenamefont {Porsev}, \citenamefont
  {Derevianko}, \citenamefont {Ovsiannikov}, \citenamefont {Pal’chikov},
  \citenamefont {Takamoto},\ and\ \citenamefont {Katori}}]{Hachisu2008}%
  \BibitemOpen
  \bibfield  {author} {\bibinfo {author} {\bibfnamefont {H.}~\bibnamefont
  {Hachisu}}, \bibinfo {author} {\bibfnamefont {K.}~\bibnamefont {Miyagishi}},
  \bibinfo {author} {\bibfnamefont {S.~G.}\ \bibnamefont {Porsev}}, \bibinfo
  {author} {\bibfnamefont {A.}~\bibnamefont {Derevianko}}, \bibinfo {author}
  {\bibfnamefont {V.~D.}\ \bibnamefont {Ovsiannikov}}, \bibinfo {author}
  {\bibfnamefont {V.~G.}\ \bibnamefont {Pal’chikov}}, \bibinfo {author}
  {\bibfnamefont {M.}~\bibnamefont {Takamoto}},\ and\ \bibinfo {author}
  {\bibfnamefont {H.}~\bibnamefont {Katori}},\ }\href
  {https://doi.org/10.1103/physrevlett.100.053001} {\bibfield  {journal}
  {\bibinfo  {journal} {Phys. Rev. Lett.}\ }\textbf {\bibinfo {volume} {100}},\
  \bibinfo {pages} {053001} (\bibinfo {year} {2008})}\BibitemShut {NoStop}%
\bibitem [{\citenamefont {Tyumenev}\ \emph {et~al.}(2016)\citenamefont
  {Tyumenev}, \citenamefont {Favier}, \citenamefont {Bilicki}, \citenamefont
  {Bookjans}, \citenamefont {Targat}, \citenamefont {Lodewyck}, \citenamefont
  {Nicolodi}, \citenamefont {Coq}, \citenamefont {Abgrall}, \citenamefont
  {Guéna}, \citenamefont {Sarlo},\ and\ \citenamefont {Bize}}]{Tyumenev2016}%
  \BibitemOpen
  \bibfield  {author} {\bibinfo {author} {\bibfnamefont {R.}~\bibnamefont
  {Tyumenev}}, \bibinfo {author} {\bibfnamefont {M.}~\bibnamefont {Favier}},
  \bibinfo {author} {\bibfnamefont {S.}~\bibnamefont {Bilicki}}, \bibinfo
  {author} {\bibfnamefont {E.}~\bibnamefont {Bookjans}}, \bibinfo {author}
  {\bibfnamefont {R.~L.}\ \bibnamefont {Targat}}, \bibinfo {author}
  {\bibfnamefont {J.}~\bibnamefont {Lodewyck}}, \bibinfo {author}
  {\bibfnamefont {D.}~\bibnamefont {Nicolodi}}, \bibinfo {author}
  {\bibfnamefont {Y.~L.}\ \bibnamefont {Coq}}, \bibinfo {author} {\bibfnamefont
  {M.}~\bibnamefont {Abgrall}}, \bibinfo {author} {\bibfnamefont
  {J.}~\bibnamefont {Guéna}}, \bibinfo {author} {\bibfnamefont {L.~D.}\
  \bibnamefont {Sarlo}},\ and\ \bibinfo {author} {\bibfnamefont
  {S.}~\bibnamefont {Bize}},\ }\href
  {https://doi.org/10.1088/1367-2630/18/11/113002} {\bibfield  {journal}
  {\bibinfo  {journal} {N. J. Phys.}\ }\textbf {\bibinfo {volume} {18}},\
  \bibinfo {pages} {113002} (\bibinfo {year} {2016})}\BibitemShut {NoStop}%
\bibitem [{\citenamefont {Schelfhout}\ and\ \citenamefont
  {McFerran}(2022)}]{Schelfhout2022}%
  \BibitemOpen
  \bibfield  {author} {\bibinfo {author} {\bibfnamefont {J.~S.}\ \bibnamefont
  {Schelfhout}}\ and\ \bibinfo {author} {\bibfnamefont {J.~J.}\ \bibnamefont
  {McFerran}},\ }\href {https://doi.org/10.1103/physreva.105.022805} {\bibfield
   {journal} {\bibinfo  {journal} {Phys. Rev. A}\ }\textbf {\bibinfo {volume}
  {105}},\ \bibinfo {pages} {022805} (\bibinfo {year} {2022})}\BibitemShut
  {NoStop}%
\bibitem [{\citenamefont {Burt}\ \emph {et~al.}(2021)\citenamefont {Burt},
  \citenamefont {Prestage}, \citenamefont {Tjoelker}, \citenamefont {Enzer},
  \citenamefont {Kuang}, \citenamefont {Murphy}, \citenamefont {Robison},
  \citenamefont {Seubert}, \citenamefont {Wang},\ and\ \citenamefont
  {Ely}}]{Burt:2021}%
  \BibitemOpen
  \bibfield  {author} {\bibinfo {author} {\bibfnamefont {E.~A.}\ \bibnamefont
  {Burt}}, \bibinfo {author} {\bibfnamefont {J.~D.}\ \bibnamefont {Prestage}},
  \bibinfo {author} {\bibfnamefont {R.~L.}\ \bibnamefont {Tjoelker}}, \bibinfo
  {author} {\bibfnamefont {D.~G.}\ \bibnamefont {Enzer}}, \bibinfo {author}
  {\bibfnamefont {D.}~\bibnamefont {Kuang}}, \bibinfo {author} {\bibfnamefont
  {D.~W.}\ \bibnamefont {Murphy}}, \bibinfo {author} {\bibfnamefont {D.~E.}\
  \bibnamefont {Robison}}, \bibinfo {author} {\bibfnamefont {J.~M.}\
  \bibnamefont {Seubert}}, \bibinfo {author} {\bibfnamefont {R.~T.}\
  \bibnamefont {Wang}},\ and\ \bibinfo {author} {\bibfnamefont {T.~A.}\
  \bibnamefont {Ely}},\ }\href {https://doi.org/10.1038/s41586-021-03571-7}
  {\bibfield  {journal} {\bibinfo  {journal} {Nature}\ }\textbf {\bibinfo
  {volume} {595}},\ \bibinfo {pages} {43} (\bibinfo {year} {2021})}\BibitemShut
  {NoStop}%
\bibitem [{\citenamefont {Sels}\ \emph {et~al.}(2019)\citenamefont {Sels},
  \citenamefont {{Day Goodacre}}, \citenamefont {Marsh}, \citenamefont
  {Pastore}, \citenamefont {Ryssens}, \citenamefont {Tsunoda}, \citenamefont
  {Althubiti}, \citenamefont {Andel}, \citenamefont {Andreyev}, \citenamefont
  {Atanasov}, \citenamefont {Barzakh}, \citenamefont {Bender}, \citenamefont
  {Billowes}, \citenamefont {Blaum}, \citenamefont {Cocolios}, \citenamefont
  {Cubiss}, \citenamefont {Dobaczewski}, \citenamefont {Farooq-Smith},
  \citenamefont {Fedorov}, \citenamefont {Fedosseev}, \citenamefont {Flanagan},
  \citenamefont {Gaffney}, \citenamefont {Ghys}, \citenamefont {Heenen},
  \citenamefont {Huyse}, \citenamefont {Kreim}, \citenamefont {Lunney},
  \citenamefont {Lynch}, \citenamefont {Manea}, \citenamefont {{Martinez
  Palenzuela}}, \citenamefont {Medonca}, \citenamefont {Molkanov},
  \citenamefont {Otsuka}, \citenamefont {Ramos}, \citenamefont {Rossel},
  \citenamefont {Rothe}, \citenamefont {Schweikhard}, \citenamefont
  {Seliverstov}, \citenamefont {Spagnoletti}, \citenamefont {{Van Beveren}},
  \citenamefont {{Van Duppen}}, \citenamefont {Veinhard}, \citenamefont
  {Verstraelen}, \citenamefont {Welker}, \citenamefont {Wendt}, \citenamefont
  {Wienholtz}, \citenamefont {Wolf},\ and\ \citenamefont
  {Zadvornaya}}]{Sels2019}%
  \BibitemOpen
  \bibfield  {author} {\bibinfo {author} {\bibfnamefont {S.}~\bibnamefont
  {Sels}}, \bibinfo {author} {\bibfnamefont {T.}~\bibnamefont {{Day
  Goodacre}}}, \bibinfo {author} {\bibfnamefont {B.~A.}\ \bibnamefont {Marsh}},
  \bibinfo {author} {\bibfnamefont {A.}~\bibnamefont {Pastore}}, \bibinfo
  {author} {\bibfnamefont {W.}~\bibnamefont {Ryssens}}, \bibinfo {author}
  {\bibfnamefont {Y.}~\bibnamefont {Tsunoda}}, \bibinfo {author} {\bibfnamefont
  {N.}~\bibnamefont {Althubiti}}, \bibinfo {author} {\bibfnamefont
  {B.}~\bibnamefont {Andel}}, \bibinfo {author} {\bibfnamefont {A.~N.}\
  \bibnamefont {Andreyev}}, \bibinfo {author} {\bibfnamefont {D.}~\bibnamefont
  {Atanasov}}, \bibinfo {author} {\bibfnamefont {A.~E.}\ \bibnamefont
  {Barzakh}}, \bibinfo {author} {\bibfnamefont {M.}~\bibnamefont {Bender}},
  \bibinfo {author} {\bibfnamefont {J.}~\bibnamefont {Billowes}}, \bibinfo
  {author} {\bibfnamefont {K.}~\bibnamefont {Blaum}}, \bibinfo {author}
  {\bibfnamefont {T.~E.}\ \bibnamefont {Cocolios}}, \bibinfo {author}
  {\bibfnamefont {J.~G.}\ \bibnamefont {Cubiss}}, \bibinfo {author}
  {\bibfnamefont {J.}~\bibnamefont {Dobaczewski}}, \bibinfo {author}
  {\bibfnamefont {G.~J.}\ \bibnamefont {Farooq-Smith}}, \bibinfo {author}
  {\bibfnamefont {D.~V.}\ \bibnamefont {Fedorov}}, \bibinfo {author}
  {\bibfnamefont {V.~N.}\ \bibnamefont {Fedosseev}}, \bibinfo {author}
  {\bibfnamefont {K.~T.}\ \bibnamefont {Flanagan}}, \bibinfo {author}
  {\bibfnamefont {L.~P.}\ \bibnamefont {Gaffney}}, \bibinfo {author}
  {\bibfnamefont {L.}~\bibnamefont {Ghys}}, \bibinfo {author} {\bibfnamefont
  {P.~H.}\ \bibnamefont {Heenen}}, \bibinfo {author} {\bibfnamefont
  {M.}~\bibnamefont {Huyse}}, \bibinfo {author} {\bibfnamefont
  {S.}~\bibnamefont {Kreim}}, \bibinfo {author} {\bibfnamefont
  {D.}~\bibnamefont {Lunney}}, \bibinfo {author} {\bibfnamefont {K.~M.}\
  \bibnamefont {Lynch}}, \bibinfo {author} {\bibfnamefont {V.}~\bibnamefont
  {Manea}}, \bibinfo {author} {\bibfnamefont {Y.}~\bibnamefont {{Martinez
  Palenzuela}}}, \bibinfo {author} {\bibfnamefont {T.~M.}\ \bibnamefont
  {Medonca}}, \bibinfo {author} {\bibfnamefont {P.~L.}\ \bibnamefont
  {Molkanov}}, \bibinfo {author} {\bibfnamefont {T.}~\bibnamefont {Otsuka}},
  \bibinfo {author} {\bibfnamefont {J.~P.}\ \bibnamefont {Ramos}}, \bibinfo
  {author} {\bibfnamefont {R.~E.}\ \bibnamefont {Rossel}}, \bibinfo {author}
  {\bibfnamefont {S.}~\bibnamefont {Rothe}}, \bibinfo {author} {\bibfnamefont
  {L.}~\bibnamefont {Schweikhard}}, \bibinfo {author} {\bibfnamefont {M.~D.}\
  \bibnamefont {Seliverstov}}, \bibinfo {author} {\bibfnamefont
  {P.}~\bibnamefont {Spagnoletti}}, \bibinfo {author} {\bibfnamefont
  {C.}~\bibnamefont {{Van Beveren}}}, \bibinfo {author} {\bibfnamefont
  {P.}~\bibnamefont {{Van Duppen}}}, \bibinfo {author} {\bibfnamefont
  {M.}~\bibnamefont {Veinhard}}, \bibinfo {author} {\bibfnamefont
  {E.}~\bibnamefont {Verstraelen}}, \bibinfo {author} {\bibfnamefont
  {A.}~\bibnamefont {Welker}}, \bibinfo {author} {\bibfnamefont
  {K.}~\bibnamefont {Wendt}}, \bibinfo {author} {\bibfnamefont
  {F.}~\bibnamefont {Wienholtz}}, \bibinfo {author} {\bibfnamefont {R.~N.}\
  \bibnamefont {Wolf}},\ and\ \bibinfo {author} {\bibfnamefont
  {A.}~\bibnamefont {Zadvornaya}},\ }\href
  {https://doi.org/10.1103/PhysRevC.99.044306} {\bibfield  {journal} {\bibinfo
  {journal} {Phys. Rev. C}\ }\textbf {\bibinfo {volume} {99}},\ \bibinfo
  {pages} {044306} (\bibinfo {year} {2019})}\BibitemShut {NoStop}%
\bibitem [{\citenamefont {Marsh}\ \emph {et~al.}(2018)\citenamefont {Marsh},
  \citenamefont {{Day Goodacre}}, \citenamefont {Sels}, \citenamefont
  {Tsunoda}, \citenamefont {Andel}, \citenamefont {Andreyev}, \citenamefont
  {Althubiti}, \citenamefont {Atanasov}, \citenamefont {Barzakh}, \citenamefont
  {Billowes}, \citenamefont {Blaum}, \citenamefont {Cocolios}, \citenamefont
  {Cubiss}, \citenamefont {Dobaczewski}, \citenamefont {Farooq-Smith},
  \citenamefont {Fedorov}, \citenamefont {Fedosseev}, \citenamefont {Flanagan},
  \citenamefont {Gaffney}, \citenamefont {Ghys}, \citenamefont {Huyse},
  \citenamefont {Kreim}, \citenamefont {Lunney}, \citenamefont {Lynch},
  \citenamefont {Manea}, \citenamefont {{Martinez Palenzuela}}, \citenamefont
  {Molkanov}, \citenamefont {Otsuka}, \citenamefont {Pastore}, \citenamefont
  {Rosenbusch}, \citenamefont {Rossel}, \citenamefont {Rothe}, \citenamefont
  {Schweikhard}, \citenamefont {Seliverstov}, \citenamefont {Spagnoletti},
  \citenamefont {{Van Beveren}}, \citenamefont {{Van Duppen}}, \citenamefont
  {Veinhard}, \citenamefont {Verstraelen}, \citenamefont {Welker},
  \citenamefont {Wendt}, \citenamefont {Wienholtz}, \citenamefont {Wolf},
  \citenamefont {Zadvornaya},\ and\ \citenamefont {Zuber}}]{Marsh2018}%
  \BibitemOpen
  \bibfield  {author} {\bibinfo {author} {\bibfnamefont {B.~A.}\ \bibnamefont
  {Marsh}}, \bibinfo {author} {\bibfnamefont {T.}~\bibnamefont {{Day
  Goodacre}}}, \bibinfo {author} {\bibfnamefont {S.}~\bibnamefont {Sels}},
  \bibinfo {author} {\bibfnamefont {Y.}~\bibnamefont {Tsunoda}}, \bibinfo
  {author} {\bibfnamefont {B.}~\bibnamefont {Andel}}, \bibinfo {author}
  {\bibfnamefont {A.~N.}\ \bibnamefont {Andreyev}}, \bibinfo {author}
  {\bibfnamefont {N.~A.}\ \bibnamefont {Althubiti}}, \bibinfo {author}
  {\bibfnamefont {D.}~\bibnamefont {Atanasov}}, \bibinfo {author}
  {\bibfnamefont {A.~E.}\ \bibnamefont {Barzakh}}, \bibinfo {author}
  {\bibfnamefont {J.}~\bibnamefont {Billowes}}, \bibinfo {author}
  {\bibfnamefont {K.}~\bibnamefont {Blaum}}, \bibinfo {author} {\bibfnamefont
  {T.~E.}\ \bibnamefont {Cocolios}}, \bibinfo {author} {\bibfnamefont {J.~G.}\
  \bibnamefont {Cubiss}}, \bibinfo {author} {\bibfnamefont {J.}~\bibnamefont
  {Dobaczewski}}, \bibinfo {author} {\bibfnamefont {G.~J.}\ \bibnamefont
  {Farooq-Smith}}, \bibinfo {author} {\bibfnamefont {D.~V.}\ \bibnamefont
  {Fedorov}}, \bibinfo {author} {\bibfnamefont {V.~N.}\ \bibnamefont
  {Fedosseev}}, \bibinfo {author} {\bibfnamefont {K.~T.}\ \bibnamefont
  {Flanagan}}, \bibinfo {author} {\bibfnamefont {L.~P.}\ \bibnamefont
  {Gaffney}}, \bibinfo {author} {\bibfnamefont {L.}~\bibnamefont {Ghys}},
  \bibinfo {author} {\bibfnamefont {M.}~\bibnamefont {Huyse}}, \bibinfo
  {author} {\bibfnamefont {S.}~\bibnamefont {Kreim}}, \bibinfo {author}
  {\bibfnamefont {D.}~\bibnamefont {Lunney}}, \bibinfo {author} {\bibfnamefont
  {K.~M.}\ \bibnamefont {Lynch}}, \bibinfo {author} {\bibfnamefont
  {V.}~\bibnamefont {Manea}}, \bibinfo {author} {\bibfnamefont
  {Y.}~\bibnamefont {{Martinez Palenzuela}}}, \bibinfo {author} {\bibfnamefont
  {P.~L.}\ \bibnamefont {Molkanov}}, \bibinfo {author} {\bibfnamefont
  {T.}~\bibnamefont {Otsuka}}, \bibinfo {author} {\bibfnamefont
  {A.}~\bibnamefont {Pastore}}, \bibinfo {author} {\bibfnamefont
  {M.}~\bibnamefont {Rosenbusch}}, \bibinfo {author} {\bibfnamefont {R.~E.}\
  \bibnamefont {Rossel}}, \bibinfo {author} {\bibfnamefont {S.}~\bibnamefont
  {Rothe}}, \bibinfo {author} {\bibfnamefont {L.}~\bibnamefont {Schweikhard}},
  \bibinfo {author} {\bibfnamefont {M.~D.}\ \bibnamefont {Seliverstov}},
  \bibinfo {author} {\bibfnamefont {P.}~\bibnamefont {Spagnoletti}}, \bibinfo
  {author} {\bibfnamefont {C.}~\bibnamefont {{Van Beveren}}}, \bibinfo {author}
  {\bibfnamefont {P.}~\bibnamefont {{Van Duppen}}}, \bibinfo {author}
  {\bibfnamefont {M.}~\bibnamefont {Veinhard}}, \bibinfo {author}
  {\bibfnamefont {E.}~\bibnamefont {Verstraelen}}, \bibinfo {author}
  {\bibfnamefont {A.}~\bibnamefont {Welker}}, \bibinfo {author} {\bibfnamefont
  {K.}~\bibnamefont {Wendt}}, \bibinfo {author} {\bibfnamefont
  {F.}~\bibnamefont {Wienholtz}}, \bibinfo {author} {\bibfnamefont {R.~N.}\
  \bibnamefont {Wolf}}, \bibinfo {author} {\bibfnamefont {A.}~\bibnamefont
  {Zadvornaya}},\ and\ \bibinfo {author} {\bibfnamefont {K.}~\bibnamefont
  {Zuber}},\ }\href {https://doi.org/10.1038/s41567-018-0292-8} {\bibfield
  {journal} {\bibinfo  {journal} {Nat. Phys.}\ }\textbf {\bibinfo {volume}
  {14}},\ \bibinfo {pages} {1163} (\bibinfo {year} {2018})}\BibitemShut
  {NoStop}%
\bibitem [{\citenamefont {Heyde}\ and\ \citenamefont {Wood}(2011)}]{Heyde2011}%
  \BibitemOpen
  \bibfield  {author} {\bibinfo {author} {\bibfnamefont {K.}~\bibnamefont
  {Heyde}}\ and\ \bibinfo {author} {\bibfnamefont {J.~L.}\ \bibnamefont
  {Wood}},\ }\href {https://doi.org/10.1103/RevModPhys.83.1467} {\bibfield
  {journal} {\bibinfo  {journal} {Rev. Mod. Phys.}\ }\textbf {\bibinfo {volume}
  {83}},\ \bibinfo {pages} {1467} (\bibinfo {year} {2011})}\BibitemShut
  {NoStop}%
\bibitem [{\citenamefont {Borie}\ and\ \citenamefont
  {Rinker}(1982)}]{Borie1982}%
  \BibitemOpen
  \bibfield  {author} {\bibinfo {author} {\bibfnamefont {E.}~\bibnamefont
  {Borie}}\ and\ \bibinfo {author} {\bibfnamefont {G.~A.}\ \bibnamefont
  {Rinker}},\ }\href {https://doi.org/10.1103/RevModPhys.54.67} {\bibfield
  {journal} {\bibinfo  {journal} {Rev. Mod. Phys.}\ }\textbf {\bibinfo {volume}
  {54}},\ \bibinfo {pages} {67} (\bibinfo {year} {1982})}\BibitemShut {NoStop}%
\bibitem [{\citenamefont {Angeli}\ and\ \citenamefont
  {Marinova}(2013)}]{Angeli2013}%
  \BibitemOpen
  \bibfield  {author} {\bibinfo {author} {\bibfnamefont {I.}~\bibnamefont
  {Angeli}}\ and\ \bibinfo {author} {\bibfnamefont {K.~P.}\ \bibnamefont
  {Marinova}},\ }\href
  {https://doi.org/https://doi.org/10.1016/j.adt.2011.12.006} {\bibfield
  {journal} {\bibinfo  {journal} {At. Data Nucl. Data Tables}\ }\textbf
  {\bibinfo {volume} {99}},\ \bibinfo {pages} {69} (\bibinfo {year}
  {2013})}\BibitemShut {NoStop}%
\bibitem [{\citenamefont {Rosenthal}\ and\ \citenamefont
  {Breit}(1932)}]{Rosenthal1932}%
  \BibitemOpen
  \bibfield  {author} {\bibinfo {author} {\bibfnamefont {J.~E.}\ \bibnamefont
  {Rosenthal}}\ and\ \bibinfo {author} {\bibfnamefont {G.}~\bibnamefont
  {Breit}},\ }\href {https://doi.org/10.1103/physrev.41.459} {\bibfield
  {journal} {\bibinfo  {journal} {Phys. Rev.}\ }\textbf {\bibinfo {volume}
  {41}},\ \bibinfo {pages} {459} (\bibinfo {year} {1932})}\BibitemShut
  {NoStop}%
\bibitem [{\citenamefont {Crawford}\ and\ \citenamefont
  {Schawlow}(1949)}]{Crawford1949}%
  \BibitemOpen
  \bibfield  {author} {\bibinfo {author} {\bibfnamefont {M.~F.}\ \bibnamefont
  {Crawford}}\ and\ \bibinfo {author} {\bibfnamefont {A.~L.}\ \bibnamefont
  {Schawlow}},\ }\href {https://doi.org/10.1103/PhysRev.76.1310} {\bibfield
  {journal} {\bibinfo  {journal} {Phys. Rev.}\ }\textbf {\bibinfo {volume}
  {76}},\ \bibinfo {pages} {1310} (\bibinfo {year} {1949})}\BibitemShut
  {NoStop}%
\bibitem [{\citenamefont {Elizarov}\ \emph {et~al.}(2006)\citenamefont
  {Elizarov}, \citenamefont {Shabaev}, \citenamefont {Oreshkina}, \citenamefont
  {Tupitsyn},\ and\ \citenamefont {St{\"{o}}hlker}}]{Elizarov2006}%
  \BibitemOpen
  \bibfield  {author} {\bibinfo {author} {\bibfnamefont {A.~A.}\ \bibnamefont
  {Elizarov}}, \bibinfo {author} {\bibfnamefont {V.~M.}\ \bibnamefont
  {Shabaev}}, \bibinfo {author} {\bibfnamefont {N.~S.}\ \bibnamefont
  {Oreshkina}}, \bibinfo {author} {\bibfnamefont {I.~I.}\ \bibnamefont
  {Tupitsyn}},\ and\ \bibinfo {author} {\bibfnamefont {T.}~\bibnamefont
  {St{\"{o}}hlker}},\ }\href {https://doi.org/10.1134/S0030400X0603009X}
  {\bibfield  {journal} {\bibinfo  {journal} {Opt. Spectrosc.}\ }\textbf
  {\bibinfo {volume} {100}},\ \bibinfo {pages} {361} (\bibinfo {year}
  {2006})}\BibitemShut {NoStop}%
\bibitem [{\citenamefont {Uehling}(1935)}]{Uehling1935}%
  \BibitemOpen
  \bibfield  {author} {\bibinfo {author} {\bibfnamefont {E.~A.}\ \bibnamefont
  {Uehling}},\ }\href {https://doi.org/10.1103/PhysRev.48.55} {\bibfield
  {journal} {\bibinfo  {journal} {Phys. Rev.}\ }\textbf {\bibinfo {volume}
  {48}},\ \bibinfo {pages} {55} (\bibinfo {year} {1935})}\BibitemShut {NoStop}%
\bibitem [{\citenamefont {Volotka}\ \emph {et~al.}(2008)\citenamefont
  {Volotka}, \citenamefont {Glazov}, \citenamefont {Tupitsyn}, \citenamefont
  {Oreshkina}, \citenamefont {Plunien},\ and\ \citenamefont
  {Shabaev}}]{Volotka2008}%
  \BibitemOpen
  \bibfield  {author} {\bibinfo {author} {\bibfnamefont {A.~V.}\ \bibnamefont
  {Volotka}}, \bibinfo {author} {\bibfnamefont {D.~A.}\ \bibnamefont {Glazov}},
  \bibinfo {author} {\bibfnamefont {I.~I.}\ \bibnamefont {Tupitsyn}}, \bibinfo
  {author} {\bibfnamefont {N.~S.}\ \bibnamefont {Oreshkina}}, \bibinfo {author}
  {\bibfnamefont {G.}~\bibnamefont {Plunien}},\ and\ \bibinfo {author}
  {\bibfnamefont {V.~M.}\ \bibnamefont {Shabaev}},\ }\href
  {https://doi.org/10.1103/PhysRevA.78.062507} {\bibfield  {journal} {\bibinfo
  {journal} {Phys. Rev. A}\ }\textbf {\bibinfo {volume} {78}},\ \bibinfo
  {pages} {062507} (\bibinfo {year} {2008})}\BibitemShut {NoStop}%
\bibitem [{\citenamefont {Shabaev}\ \emph {et~al.}(2001)\citenamefont
  {Shabaev}, \citenamefont {Artemyev}, \citenamefont {Yerokhin}, \citenamefont
  {Zherebtsov},\ and\ \citenamefont {Soff}}]{Shabaev2001}%
  \BibitemOpen
  \bibfield  {author} {\bibinfo {author} {\bibfnamefont {V.~M.}\ \bibnamefont
  {Shabaev}}, \bibinfo {author} {\bibfnamefont {A.~N.}\ \bibnamefont
  {Artemyev}}, \bibinfo {author} {\bibfnamefont {V.~A.}\ \bibnamefont
  {Yerokhin}}, \bibinfo {author} {\bibfnamefont {O.~M.}\ \bibnamefont
  {Zherebtsov}},\ and\ \bibinfo {author} {\bibfnamefont {G.}~\bibnamefont
  {Soff}},\ }\href {https://doi.org/10.1103/PhysRevLett.86.3959} {\bibfield
  {journal} {\bibinfo  {journal} {Phys. Rev. Lett.}\ }\textbf {\bibinfo
  {volume} {86}},\ \bibinfo {pages} {3959} (\bibinfo {year}
  {2001})}\BibitemShut {NoStop}%
\bibitem [{\citenamefont {Persson}(2023)}]{Persson2023}%
  \BibitemOpen
  \bibfield  {author} {\bibinfo {author} {\bibfnamefont {J.~R.}\ \bibnamefont
  {Persson}},\ }\href {https://doi.org/10.1016/j.adt.2023.101589} {\bibfield
  {journal} {\bibinfo  {journal} {Atomic Data and Nuclear Data Tables}\
  }\textbf {\bibinfo {volume} {154}},\ \bibinfo {pages} {101589} (\bibinfo
  {year} {2023})}\BibitemShut {NoStop}%
\bibitem [{\citenamefont {Reimann}\ and\ \citenamefont
  {McDermott}(1973)}]{Reimann1973}%
  \BibitemOpen
  \bibfield  {author} {\bibinfo {author} {\bibfnamefont {R.~J.}\ \bibnamefont
  {Reimann}}\ and\ \bibinfo {author} {\bibfnamefont {M.~N.}\ \bibnamefont
  {McDermott}},\ }\href {https://doi.org/10.1103/physrevc.7.2065} {\bibfield
  {journal} {\bibinfo  {journal} {Phys. Rev. C}\ }\textbf {\bibinfo {volume}
  {7}},\ \bibinfo {pages} {2065} (\bibinfo {year} {1973})}\BibitemShut
  {NoStop}%
\bibitem [{\citenamefont {Grunefeld}\ \emph {et~al.}(2019)\citenamefont
  {Grunefeld}, \citenamefont {Roberts},\ and\ \citenamefont
  {Ginges}}]{Grunefeld2019}%
  \BibitemOpen
  \bibfield  {author} {\bibinfo {author} {\bibfnamefont {S.~J.}\ \bibnamefont
  {Grunefeld}}, \bibinfo {author} {\bibfnamefont {B.~M.}\ \bibnamefont
  {Roberts}},\ and\ \bibinfo {author} {\bibfnamefont {J.~S.~M.}\ \bibnamefont
  {Ginges}},\ }\href {https://doi.org/10.1103/PhysRevA.100.042506} {\bibfield
  {journal} {\bibinfo  {journal} {Phys. Rev. A}\ }\textbf {\bibinfo {volume}
  {100}},\ \bibinfo {pages} {042506} (\bibinfo {year} {2019})}\BibitemShut
  {NoStop}%
\bibitem [{\citenamefont {Roberts}\ \emph {et~al.}(2023)\citenamefont
  {Roberts}, \citenamefont {Fairhall},\ and\ \citenamefont
  {Ginges}}]{Roberts2022}%
  \BibitemOpen
  \bibfield  {author} {\bibinfo {author} {\bibfnamefont {B.~M.}\ \bibnamefont
  {Roberts}}, \bibinfo {author} {\bibfnamefont {C.~J.}\ \bibnamefont
  {Fairhall}},\ and\ \bibinfo {author} {\bibfnamefont {J.~S.~M.}\ \bibnamefont
  {Ginges}},\ }\href {https://doi.org/PhysRevA.107.052812} {\bibfield
  {journal} {\bibinfo  {journal} {Phys. Rev. A}\ }\textbf {\bibinfo {volume}
  {107}},\ \bibinfo {pages} {052812} (\bibinfo {year} {2023})}\BibitemShut
  {NoStop}%
\bibitem [{\citenamefont {Dzuba}\ \emph {et~al.}(1988)\citenamefont {Dzuba},
  \citenamefont {Flambaum}, \citenamefont {Silvestrov},\ and\ \citenamefont
  {Sushkov}}]{Dzuba1988}%
  \BibitemOpen
  \bibfield  {author} {\bibinfo {author} {\bibfnamefont {V.~A.}\ \bibnamefont
  {Dzuba}}, \bibinfo {author} {\bibfnamefont {V.~V.}\ \bibnamefont {Flambaum}},
  \bibinfo {author} {\bibfnamefont {P.~G.}\ \bibnamefont {Silvestrov}},\ and\
  \bibinfo {author} {\bibfnamefont {O.~P.}\ \bibnamefont {Sushkov}},\ }\href
  {https://doi.org/https://doi.org/10.1016/0375-9601(88)90302-7} {\bibfield
  {journal} {\bibinfo  {journal} {Phys. Lett. A}\ }\textbf {\bibinfo {volume}
  {131}},\ \bibinfo {pages} {461} (\bibinfo {year} {1988})}\BibitemShut
  {NoStop}%
\bibitem [{\citenamefont {Stone}(2019)}]{Stone2019}%
  \BibitemOpen
  \bibfield  {author} {\bibinfo {author} {\bibfnamefont {N.~J.}\ \bibnamefont
  {Stone}},\ }\href {https://doi.org/10.61092/iaea.yjpc-cns6} {\emph {\bibinfo
  {title} {Table of Recommended Nuclear Magnetic Dipole Moments}}}\ (\bibinfo
  {publisher} {International Atomic Energy Agency (IAEA)},\ \bibinfo {year}
  {2019})\BibitemShut {NoStop}%
\bibitem [{\citenamefont {Ulm}\ \emph {et~al.}(1986)\citenamefont {Ulm},
  \citenamefont {Bhattacherjee}, \citenamefont {Dabkiewicz}, \citenamefont
  {Huber}, \citenamefont {Kluge}, \citenamefont {K{\"{u}}hl}, \citenamefont
  {Lochmann}, \citenamefont {Otten}, \citenamefont {Wendt}, \citenamefont
  {Ahmad}, \citenamefont {Klempt},\ and\ \citenamefont {Neugart}}]{Ulm1986}%
  \BibitemOpen
  \bibfield  {author} {\bibinfo {author} {\bibfnamefont {G.}~\bibnamefont
  {Ulm}}, \bibinfo {author} {\bibfnamefont {S.~K.}\ \bibnamefont
  {Bhattacherjee}}, \bibinfo {author} {\bibfnamefont {P.}~\bibnamefont
  {Dabkiewicz}}, \bibinfo {author} {\bibfnamefont {G.}~\bibnamefont {Huber}},
  \bibinfo {author} {\bibfnamefont {H.~J.}\ \bibnamefont {Kluge}}, \bibinfo
  {author} {\bibfnamefont {T.}~\bibnamefont {K{\"{u}}hl}}, \bibinfo {author}
  {\bibfnamefont {H.}~\bibnamefont {Lochmann}}, \bibinfo {author}
  {\bibfnamefont {E.~W.}\ \bibnamefont {Otten}}, \bibinfo {author}
  {\bibfnamefont {K.}~\bibnamefont {Wendt}}, \bibinfo {author} {\bibfnamefont
  {S.~A.}\ \bibnamefont {Ahmad}}, \bibinfo {author} {\bibfnamefont
  {W.}~\bibnamefont {Klempt}},\ and\ \bibinfo {author} {\bibfnamefont
  {R.}~\bibnamefont {Neugart}},\ }\href {https://doi.org/10.1007/BF01294605}
  {\bibfield  {journal} {\bibinfo  {journal} {Z. Phys. A}\ }\textbf {\bibinfo
  {volume} {325}},\ \bibinfo {pages} {247} (\bibinfo {year}
  {1986})}\BibitemShut {NoStop}%
\bibitem [{\citenamefont {Burt}\ \emph {et~al.}(2009)\citenamefont {Burt},
  \citenamefont {Taghavi-Larigani},\ and\ \citenamefont {Tjoelker}}]{Burt2009}%
  \BibitemOpen
  \bibfield  {author} {\bibinfo {author} {\bibfnamefont {E.~A.}\ \bibnamefont
  {Burt}}, \bibinfo {author} {\bibfnamefont {S.}~\bibnamefont
  {Taghavi-Larigani}},\ and\ \bibinfo {author} {\bibfnamefont {R.~L.}\
  \bibnamefont {Tjoelker}},\ }\href
  {https://doi.org/10.1103/PhysRevA.79.062506} {\bibfield  {journal} {\bibinfo
  {journal} {Phys. Rev. A}\ }\textbf {\bibinfo {volume} {79}},\ \bibinfo
  {pages} {062506} (\bibinfo {year} {2009})}\BibitemShut {NoStop}%
\bibitem [{\citenamefont {Antognini}\ \emph {et~al.}(2020)\citenamefont
  {Antognini}, \citenamefont {Berger}, \citenamefont {Cocolios}, \citenamefont
  {Dressler}, \citenamefont {Eichler}, \citenamefont {Eggenberger},
  \citenamefont {Indelicato}, \citenamefont {Jungmann}, \citenamefont {Keitel},
  \citenamefont {Kirch}, \citenamefont {Knecht}, \citenamefont {Michel},
  \citenamefont {Nuber}, \citenamefont {Oreshkina}, \citenamefont {Ouf},
  \citenamefont {Papa}, \citenamefont {Pohl}, \citenamefont {Pospelov},
  \citenamefont {Rapisarda}, \citenamefont {Ritjoho}, \citenamefont {Roccia},
  \citenamefont {Severijns}, \citenamefont {Skawran}, \citenamefont {Vogiatzi},
  \citenamefont {Wauters},\ and\ \citenamefont {Willmann}}]{Antognini2020}%
  \BibitemOpen
  \bibfield  {author} {\bibinfo {author} {\bibfnamefont {A.}~\bibnamefont
  {Antognini}}, \bibinfo {author} {\bibfnamefont {N.}~\bibnamefont {Berger}},
  \bibinfo {author} {\bibfnamefont {T.~E.}\ \bibnamefont {Cocolios}}, \bibinfo
  {author} {\bibfnamefont {R.}~\bibnamefont {Dressler}}, \bibinfo {author}
  {\bibfnamefont {R.}~\bibnamefont {Eichler}}, \bibinfo {author} {\bibfnamefont
  {A.}~\bibnamefont {Eggenberger}}, \bibinfo {author} {\bibfnamefont
  {P.}~\bibnamefont {Indelicato}}, \bibinfo {author} {\bibfnamefont
  {K.}~\bibnamefont {Jungmann}}, \bibinfo {author} {\bibfnamefont {C.~H.}\
  \bibnamefont {Keitel}}, \bibinfo {author} {\bibfnamefont {K.}~\bibnamefont
  {Kirch}}, \bibinfo {author} {\bibfnamefont {A.}~\bibnamefont {Knecht}},
  \bibinfo {author} {\bibfnamefont {N.}~\bibnamefont {Michel}}, \bibinfo
  {author} {\bibfnamefont {J.}~\bibnamefont {Nuber}}, \bibinfo {author}
  {\bibfnamefont {N.~S.}\ \bibnamefont {Oreshkina}}, \bibinfo {author}
  {\bibfnamefont {A.}~\bibnamefont {Ouf}}, \bibinfo {author} {\bibfnamefont
  {A.}~\bibnamefont {Papa}}, \bibinfo {author} {\bibfnamefont {R.}~\bibnamefont
  {Pohl}}, \bibinfo {author} {\bibfnamefont {M.}~\bibnamefont {Pospelov}},
  \bibinfo {author} {\bibfnamefont {E.}~\bibnamefont {Rapisarda}}, \bibinfo
  {author} {\bibfnamefont {N.}~\bibnamefont {Ritjoho}}, \bibinfo {author}
  {\bibfnamefont {S.}~\bibnamefont {Roccia}}, \bibinfo {author} {\bibfnamefont
  {N.}~\bibnamefont {Severijns}}, \bibinfo {author} {\bibfnamefont
  {A.}~\bibnamefont {Skawran}}, \bibinfo {author} {\bibfnamefont {S.~M.}\
  \bibnamefont {Vogiatzi}}, \bibinfo {author} {\bibfnamefont {F.}~\bibnamefont
  {Wauters}},\ and\ \bibinfo {author} {\bibfnamefont {L.}~\bibnamefont
  {Willmann}},\ }\href {https://doi.org/10.1103/PhysRevC.101.054313} {\bibfield
   {journal} {\bibinfo  {journal} {Phys. Rev. C}\ }\textbf {\bibinfo {volume}
  {101}},\ \bibinfo {pages} {054313} (\bibinfo {year} {2020})}\BibitemShut
  {NoStop}%
\bibitem [{\citenamefont {Wauters}\ \emph {et~al.}(2021)\citenamefont
  {Wauters}, \citenamefont {Knecht},\ and\ \citenamefont {muX
  collaboration}}]{Wauters2021}%
  \BibitemOpen
  \bibfield  {author} {\bibinfo {author} {\bibfnamefont {F.}~\bibnamefont
  {Wauters}}, \bibinfo {author} {\bibfnamefont {A.}~\bibnamefont {Knecht}},\
  and\ \bibinfo {author} {\bibnamefont {muX collaboration}},\ }\href
  {https://doi.org/10.21468/SciPostPhysProc.5.022} {\bibfield  {journal}
  {\bibinfo  {journal} {SciPost Phys. Proc.}\ ,\ \bibinfo {pages} {022}}
  (\bibinfo {year} {2021})}\BibitemShut {NoStop}%
\bibitem [{\citenamefont {Ullmann}\ \emph {et~al.}(2017)\citenamefont
  {Ullmann}, \citenamefont {Andelkovic}, \citenamefont {Brandau}, \citenamefont
  {Dax}, \citenamefont {Geithner}, \citenamefont {Geppert}, \citenamefont
  {Gorges}, \citenamefont {Hammen}, \citenamefont {Hannen}, \citenamefont
  {Kaufmann}, \citenamefont {K{\"{o}}nig}, \citenamefont {Litvinov},
  \citenamefont {Lochmann}, \citenamefont {Maa{\ss}}, \citenamefont {Meisner},
  \citenamefont {Murb{\"{o}}ck}, \citenamefont {S{\'{a}}nchez}, \citenamefont
  {Schmidt}, \citenamefont {Schmidt}, \citenamefont {Steck}, \citenamefont
  {St{\"{o}}hlker}, \citenamefont {Thompson}, \citenamefont {Trageser},
  \citenamefont {Vollbrecht}, \citenamefont {Weinheimer},\ and\ \citenamefont
  {N{\"{o}}rtersha{\"{u}}ser}}]{Ullmann2017}%
  \BibitemOpen
  \bibfield  {author} {\bibinfo {author} {\bibfnamefont {J.}~\bibnamefont
  {Ullmann}}, \bibinfo {author} {\bibfnamefont {Z.}~\bibnamefont {Andelkovic}},
  \bibinfo {author} {\bibfnamefont {C.}~\bibnamefont {Brandau}}, \bibinfo
  {author} {\bibfnamefont {A.}~\bibnamefont {Dax}}, \bibinfo {author}
  {\bibfnamefont {W.}~\bibnamefont {Geithner}}, \bibinfo {author}
  {\bibfnamefont {C.}~\bibnamefont {Geppert}}, \bibinfo {author} {\bibfnamefont
  {C.}~\bibnamefont {Gorges}}, \bibinfo {author} {\bibfnamefont
  {M.}~\bibnamefont {Hammen}}, \bibinfo {author} {\bibfnamefont
  {V.}~\bibnamefont {Hannen}}, \bibinfo {author} {\bibfnamefont
  {S.}~\bibnamefont {Kaufmann}}, \bibinfo {author} {\bibfnamefont
  {K.}~\bibnamefont {K{\"{o}}nig}}, \bibinfo {author} {\bibfnamefont {Y.~A.}\
  \bibnamefont {Litvinov}}, \bibinfo {author} {\bibfnamefont {M.}~\bibnamefont
  {Lochmann}}, \bibinfo {author} {\bibfnamefont {B.}~\bibnamefont {Maa{\ss}}},
  \bibinfo {author} {\bibfnamefont {J.}~\bibnamefont {Meisner}}, \bibinfo
  {author} {\bibfnamefont {T.}~\bibnamefont {Murb{\"{o}}ck}}, \bibinfo {author}
  {\bibfnamefont {R.}~\bibnamefont {S{\'{a}}nchez}}, \bibinfo {author}
  {\bibfnamefont {M.}~\bibnamefont {Schmidt}}, \bibinfo {author} {\bibfnamefont
  {S.}~\bibnamefont {Schmidt}}, \bibinfo {author} {\bibfnamefont
  {M.}~\bibnamefont {Steck}}, \bibinfo {author} {\bibfnamefont
  {T.}~\bibnamefont {St{\"{o}}hlker}}, \bibinfo {author} {\bibfnamefont
  {R.~C.}\ \bibnamefont {Thompson}}, \bibinfo {author} {\bibfnamefont
  {C.}~\bibnamefont {Trageser}}, \bibinfo {author} {\bibfnamefont
  {J.}~\bibnamefont {Vollbrecht}}, \bibinfo {author} {\bibfnamefont
  {C.}~\bibnamefont {Weinheimer}},\ and\ \bibinfo {author} {\bibfnamefont
  {W.}~\bibnamefont {N{\"{o}}rtersha{\"{u}}ser}},\ }\href
  {https://doi.org/10.1038/ncomms15484} {\bibfield  {journal} {\bibinfo
  {journal} {Nat. Commun.}\ }\textbf {\bibinfo {volume} {8}},\ \bibinfo {pages}
  {15484} (\bibinfo {year} {2017})}\BibitemShut {NoStop}%
\bibitem [{\citenamefont {Dickopf}\ \emph {et~al.}(2024)\citenamefont
  {Dickopf}, \citenamefont {Sikora}, \citenamefont {Kaiser}, \citenamefont
  {M{\"u}ller}, \citenamefont {Ulmer}, \citenamefont {Yerokhin}, \citenamefont
  {Harman}, \citenamefont {Keitel}, \citenamefont {Mooser},\ and\ \citenamefont
  {Blaum}}]{Dickopf2024}%
  \BibitemOpen
  \bibfield  {author} {\bibinfo {author} {\bibfnamefont {S.}~\bibnamefont
  {Dickopf}}, \bibinfo {author} {\bibfnamefont {B.}~\bibnamefont {Sikora}},
  \bibinfo {author} {\bibfnamefont {A.}~\bibnamefont {Kaiser}}, \bibinfo
  {author} {\bibfnamefont {M.}~\bibnamefont {M{\"u}ller}}, \bibinfo {author}
  {\bibfnamefont {S.}~\bibnamefont {Ulmer}}, \bibinfo {author} {\bibfnamefont
  {V.~A.}\ \bibnamefont {Yerokhin}}, \bibinfo {author} {\bibfnamefont
  {Z.}~\bibnamefont {Harman}}, \bibinfo {author} {\bibfnamefont {C.~H.}\
  \bibnamefont {Keitel}}, \bibinfo {author} {\bibfnamefont {A.}~\bibnamefont
  {Mooser}},\ and\ \bibinfo {author} {\bibfnamefont {K.}~\bibnamefont
  {Blaum}},\ }\href {https://doi.org/10.1038/s41586-024-07795-1} {\bibfield
  {journal} {\bibinfo  {journal} {Nature}\ }\textbf {\bibinfo {volume} {632}},\
  \bibinfo {pages} {757} (\bibinfo {year} {2024})}\BibitemShut {NoStop}%
\end{thebibliography}%
\end{document}